\PassOptionsToPackage{svgnames,prologue}{xcolor}
\documentclass[manuscript,screen,]{acmart}
\makeatletter
\AtBeginDocument{%
  \let\baselinestretch\ACM@origbaselinestretch
}
\makeatother

\AtBeginDocument{%
  }

\setcopyright{none}

\acmJournal{TOSEM}

\usepackage{lmodern}
\usepackage{tikz}
\usetikzlibrary{
  shapes,               
  arrows.meta,          
  positioning,          
  decorations.pathreplacing, 
  fit                   
}

\usepackage[all]{hypcap}
\usepackage[svgnames]{xcolor}

\usepackage{hyperref}
\hypersetup{
    colorlinks = true,
    citecolor = {blue},
}

\usepackage{algorithm}
\usepackage{algorithmicx}
\usepackage{algpseudocode}
\usepackage{caption} 
\usepackage{microtype}
\usepackage{graphicx}
\expandafter\def\csname ver@subfig.sty\endcsname{}
\usepackage{booktabs}
\usepackage{float}
\usepackage{bigstrut}
\usepackage{amsmath}

\usepackage{amssymb}
\usepackage{mathtools}
\usepackage{amsthm}
\usepackage{mathrsfs}
\usepackage{nicefrac}
\usepackage{dsfont}
\usepackage{enumitem}
\usepackage{subcaption}
\usepackage{graphicx,subfig}
\usepackage{cleveref}
\usepackage{epigraph}
\usepackage{capt-of}
\usepackage{arydshln}
\usepackage[utf8]{inputenc}
\usepackage[T1]{fontenc}
\usepackage{url}
\usepackage{amsfonts}
\usepackage{fdsymbol}
\usepackage{wrapfig}
\usepackage{lipsum}
\usepackage{stackengine}
\usepackage[font=small,labelfont=bf]{caption}
\usepackage{color}
\usepackage{adjustbox}
\usepackage{rotating}
\usepackage{makecell}
\usepackage{multirow}
\usepackage{tabularx}
\newcolumntype{Y}{>{\centering\arraybackslash}X}

\usepackage{bera}
\usepackage{listings}

\definecolor{eclipseStrings}{RGB}{42,0,255}
\definecolor{eclipseKeywords}{RGB}{127,0,85}
\colorlet{numb}{magenta!60!black}
\definecolor{bggray}{rgb}{0.95, 0.95, 0.95}
\definecolor{linenobg}{rgb}{1,1,1}

\lstdefinelanguage{json}{
    basicstyle=\footnotesize\ttfamily,
    commentstyle=\color{eclipseStrings},
    stringstyle=\color{eclipseKeywords},
    numbers=left,
    numberstyle=\scriptsize\color{black}\setlength{\fboxsep}{0pt}\colorbox{linenobg},
    stepnumber=1,
    numbersep=10pt,
    xleftmargin=2em,
    framexleftmargin=2em,
    showstringspaces=false,
    breaklines=true,
    frame=single,
    backgroundcolor=\color{bggray},
    string=[s]{"}{"},
    comment=[l]{:},
    morecomment=[l]{,},
}

\definecolor{dockerbg}{rgb}{0.97, 0.97, 0.97}
\definecolor{scriptbg}{rgb}{0.90, 0.95, 1.00}

\usepackage[most]{tcolorbox}

\tcbset{
  images/.style={
    colback=dockerbg,
    colframe=gray,
    boxrule=1pt,
    arc=0mm,
    colbacktitle=gray,
    coltitle=white,
    fonttitle=\bfseries\sffamily,
    title={TITLE},
    toptitle=2mm,
    bottomtitle=2mm,
    leftrule=2pt, rightrule=2pt,
    top=1mm, bottom=1mm, left=1mm, right=1mm,
  },
  shell/.style={
    colback=blue!5,
    colframe=blue,
    boxrule=1pt,
    arc=2mm,
    colbacktitle=blue,
    coltitle=white,
    fonttitle=\bfseries\sffamily,
    title={TITLE},
    toptitle=2mm,
    bottomtitle=2mm,
    leftrule=2pt, rightrule=2pt,
    top=1mm, bottom=1mm, left=3mm, right=3mm,
  }
}

\setlist[itemize]{topsep=0.3em, partopsep=0pt, parsep=0pt, itemsep=0.4em}

\definecolor{blanchedalmond}{rgb}{1.0, 0.92, 0.8}
\definecolor{carmine}{rgb}{0.59, 0.0, 0.09}
\definecolor{lightblue}{rgb}{0.22,0.45,0.70}%

\renewcommand{\mathbf}{\boldsymbol}

\makeatletter
\def\Ddots{\mathinner{\mkern1mu\raise\p@
\vbox{\kern7\p@\hbox{.}}\mkern2mu
\raise4\p@\hbox{.}\mkern2mu\raise7\p@\hbox{.}\mkern1mu}}
\makeatother

\definecolor{amaranth}{rgb}{0.9, 0.17, 0.31}
\definecolor{antiquebrass}{rgb}{0.8, 0.58, 0.46}
\definecolor{antiquefuchsia}{rgb}{0.57, 0.36, 0.51}
\definecolor{chromeyellow}{rgb}{0.31, 0.47, 0.26}

\newtcolorbox{AIbox}[2][]{aibox,title=#2,#1}
\definecolor{lightblue}{rgb}{0.22,0.45,0.70}%
\definecolor{Gray}{gray}{0.95}
\definecolor{Cornsilk}{rgb}{1.0, 0.97, 0.86}

\usepackage{siunitx}
\title{RepoForge: Training a SOTA Fast-thinking SWE Agent with an End-to-End Data Curation Pipeline Synergizing SFT and RL at Scale}

\author{Zhilong Chen}
\email{zhilong.chen@huawei.com}
\affiliation{%
  \institution{Center for Software Excellence, Huawei Canada}
  \country{Canada}
}
\author{Chengzong Zhao}
\email{chengzong.zhao@uwaterloo.ca}
\affiliation{%
  \institution{University of Waterloo}
  \city{Waterloo}
  \state{Ontario}
  \country{Canada}
}

\author{Boyuan Chen}
\email{boyuan.chen@huawei.com}
\affiliation{%
  \institution{Center for Software Excellence, Huawei Canada}
  \country{Canada}
}

\author{Dayi Lin}
\email{dayi.lin@huawei.com}
\affiliation{%
  \institution{Center for Software Excellence, Huawei Canada}
  \country{Canada}
}

\author{Yihao Chen}
\email{yihao.chen@huawei.com}
\affiliation{%
  \institution{Center for Software Excellence, Huawei Canada}
  \country{Canada}
}

\author{Arthur Leung}
\email{arthur.leung@huawei.com}
\affiliation{%
  \institution{Center for Software Excellence, Huawei Canada}
  \country{Canada}
}

\author{Gopi Krishnan Rajbahadur}
\email{gopi.rajbahadur@huawei.com}
\affiliation{%
  \institution{Center for Software Excellence, Huawei Canada}
  \country{Canada}
}

\author{Gustavo Oliva}
\email{gustavo.oliva@huawei.com}
\affiliation{%
  \institution{Center for Software Excellence, Huawei Canada}
  \country{Canada}
}

\author{Haoxiang Zhang}
\email{haoxiang.zhang@huawei.com}
\affiliation{%
  \institution{Center for Software Excellence, Huawei Canada}
  \country{Canada}
}

\author{Aaditya Bhatia}
\email{aaditya.bhatia@huawei.com}
\affiliation{%
  \institution{Center for Software Excellence, Huawei Canada}
  \country{Canada}
}

\author{Chong Chun Yong}
\email{chong.chun.yong@huawei.com}
\affiliation{%
  \institution{Software Engineering Laboratory, Huawei Hong Kong}
  \country{China}
}

\author{Ahmed Hassan}
\email{ahmed@cs.queensu.ca}
\affiliation{%
  \institution{Queen’s University}
  \city{Kingston}
  \country{Canada}
}



\begin{document}

\begin{abstract}
Training software engineering (SWE) LLMs is bottlenecked by expensive infrastructure, inefficient evaluation pipelines, scarce training data, and costly quality control. We present RepoForge, an autonomous, end-to-end pipeline that generates, evaluates, and trains SWE agents at scale. Our key contributions include: (1) RepoForge-8B-Agent, achieving 17.4\% on SWE-Bench-Verified~\citep{swebench_verified2024}, establishing new state-of-the-art for $\leq$8B non-thinking LLMs; (2) 7,304 executable environments auto-generated from real GitHub commits with zero manual intervention; (3) 14$\times$ storage reduction (1.4GB $\rightarrow$ 102MB per instance) via intelligent dependency management and image pruning; (4) $>$70\% faster evaluation using a Ray-powered~\citep{ray2018} distributed RepoForge harness; (5) 19,000$\times$ cheaper labeling through our automated SPICE~\citep{spice2024} difficulty assessment technique. By unifying storage-efficient sandboxing, Ray-powered evaluation harness, automated data generation, SPICE-based labeling, and bubble-free RL scaffold, we demonstrate that even $\leq$8B models can reach new state-of-the-art performance on demanding benchmarks like SWE-Bench-Verified. Our approach addresses critical bottlenecks in SWE agent training: high storage costs of container-based evaluation, inefficient sequential reward pipelines, limited availability of high-quality training data, expensive manual labeling, and multi-turn RL pipeline bottlenecks.
\end{abstract} 

\maketitle

\section{Introduction}
\label{sec:intro}

\begin{figure}[htbp]
\centering
\includegraphics[width=0.8\linewidth]{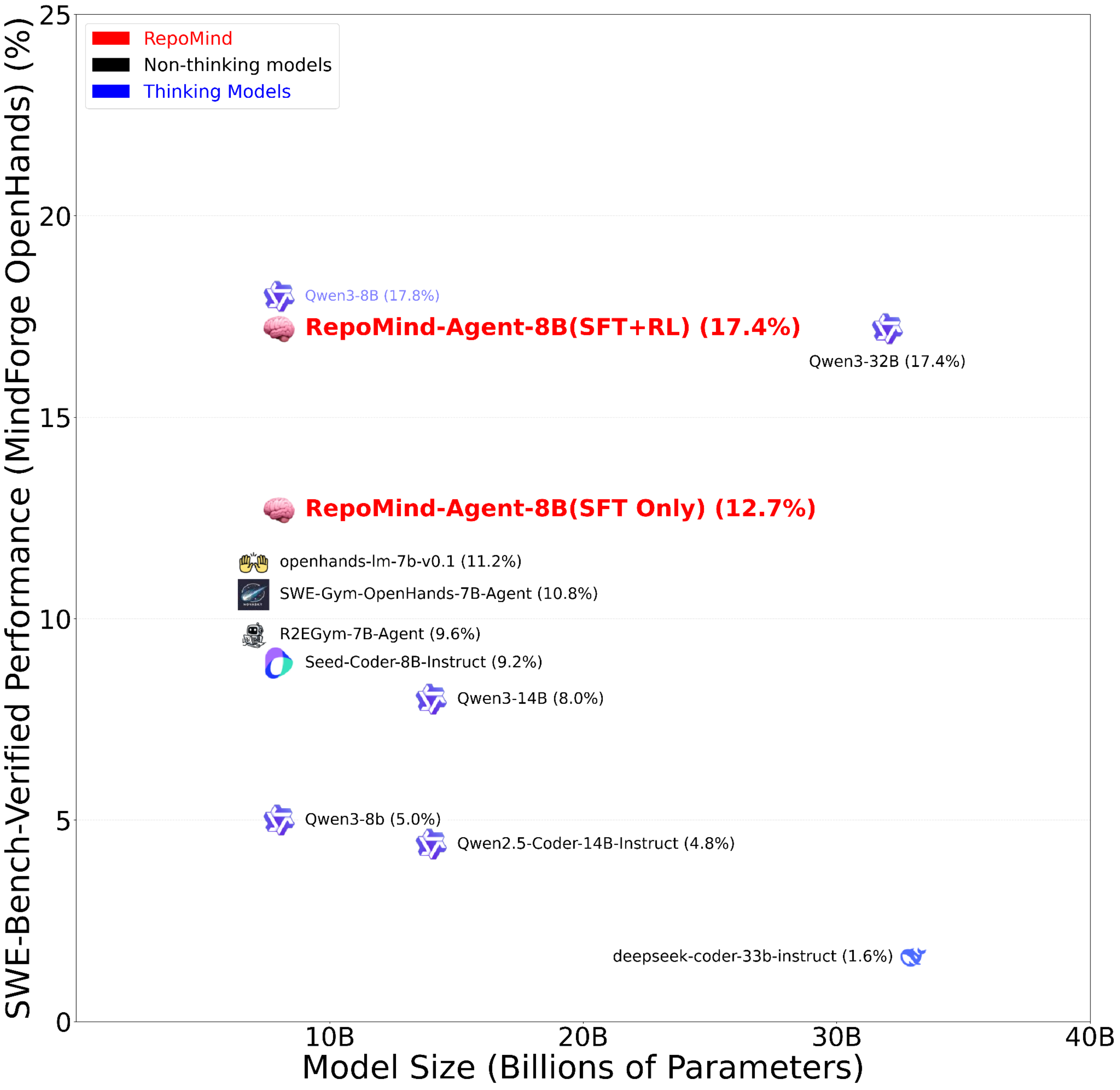}
\caption{SWE-Bench Verified Performance vs. Model Size for LLM Coding Agents with RepoForge Harness tested in-house with RepoForge OpenHands (Our improved OpenHands). RepoMind-Agent-8B, a non-thinking model, achieves 17.4\%, nearly matching the highest performing model in 8B scale.}
\label{fig:performance_RepoForge}
\end{figure} 
\begin{figure}[htbp]
\centering
\includegraphics[width=0.8\linewidth]{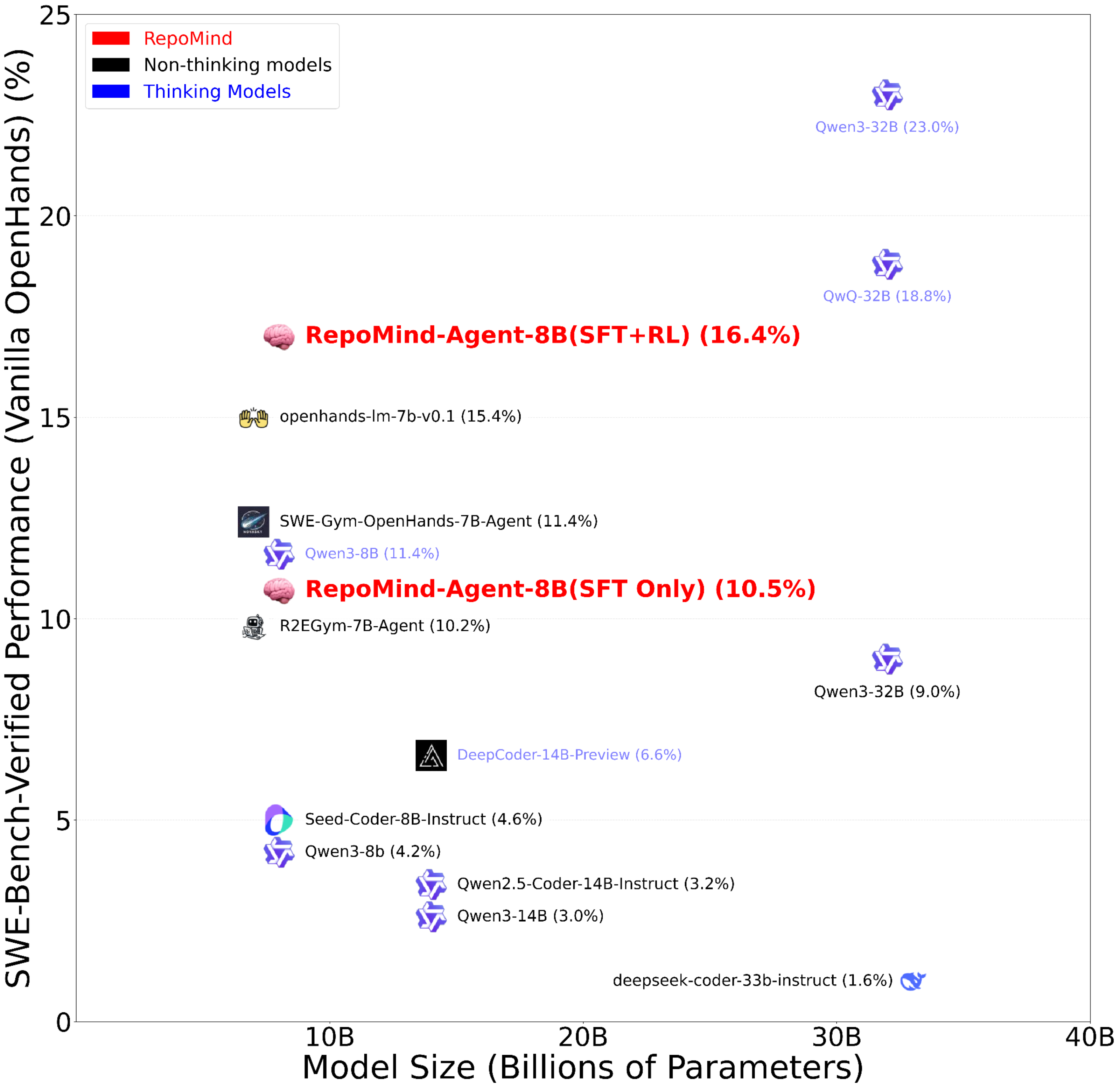}
\caption{SWE-Bench Verified Performance vs. Model Size for LLM Coding Agents with the official OpenHands tested in-house. RepoMind-Agent-8B stands out as the best performer, achieving 16.4\%.}
\label{fig:performance_official}
\end{figure} 
The last twelve months have seen language models evolve from passive inferrers into \textbf{active agents}. Early open-source frameworks such as \textbf{Search-R}1~\citep{searchr1_2024} and \textbf{MMSearch-R1}~\citep{mmsearch_r1_2024} taught models to interleave chain-of-thought with single tool invocations for web search. \textbf{ToRL}~\citep{torl2024} pushed further by scaling reinforcement learning directly from base checkpoints with unrestricted exploration, while the underlying library \textbf{veRL}~\citep{verl2024} supplied fast, efficient RLHF primitives. Most recently, \textbf{SkyRL}~\citep{skyrl2024} demonstrated that the same ideas can be extended to long-horizon environments like SWE-Bench by layering veRL with the OpenHands~\citep{openhands2024} execution scaffold.

The pace of progress in agentic code LLMs has revolutionized software engineering, automating everything from pull request drafting to full module scaffolding. But moving from benchmark demos like SkyRL and OpenHands to real-world deployment uncovers bottlenecks at every stage. The heart of these benchmarks is an evaluation harness that must spin up isolated Docker environments, install dependencies, apply patches, run tests and record pass/fail results—essentially an automated CI/CD system that validates each proposed fix through real code execution.

However, building production-scale SWE agents requires addressing challenges that extend far beyond just the evaluation harness, touching every aspect of the training pipeline. These challenges create fundamental barriers to scaling SWE agent training and deployment in real-world scenarios.

\begin{itemize}
  \item \textbf{Challenge 1: High Storage Costs of Container-Based Sandbox Evaluation} - Evaluating software engineering tasks typically requires isolating each task within dedicated Docker images (1.4 GB per image, plus approximately 7 GB per running container). This leads to significant disk usage, especially pronounced during reinforcement learning processes, which frequently execute tasks in multiple parallel rollouts over numerous training iterations. For example, evaluating merely 500 SWE-Bench-Verified like tasks demands around 690 GB of storage, easily exceeding 1 TB as RL evaluation scales up with repeated task execution, where each task requires a separate running container.
  \item \textbf{Challenge 2: Inefficient and Sequential Reward Evaluation Pipelines} - Existing evaluation framework harnesses use sequential, blocking pipelines with no artifact reuse, creating massive latency and wasted compute during training loops. 
  \item \textbf{Challenge 3: Limited Availability of High-Quality SWE Training Data} - Executable SWE training data is scarce and typically requires extensive manual effort for curation, especially for historical issues. Developers must painstakingly reconstruct and validate dependencies to ensure accurate reproduction of issue-resolution processes.
  \item  \textbf{Challenge 4: High Cost and Subjectivity of Manual Data Labeling} - Manual difficulty assessment and quality filtering are expensive, subjective, and don't scale. Expert annotation can cost \$15,000+ per 1,000 instances.
  \item \textbf{Challenge 5: Agentic Multi-Turn Reinforcement Learning Pipeline Bottlenecks} - Multi-turn reinforcement learning suffers from "pipeline bubbles" where stages stall, especially problematic for 20-50 iteration trajectories required by modern scaffolds.
\end{itemize}

RepoForge tackles all of these with a fully autonomous, end-to-end pipeline. By combining Image Dependency Pruning and Minimal Runtime Environments, it cuts per-task storage \textbf{14× (from 1.4 GB down to 0.102 GB)}. A \textbf{Ray-powered}, streaming evaluation harness drives over \textbf{70\% latency reduction}. The RepoForge Foundry autonomously generates \textbf{7,304} validated environments from real GitHub commits with zero manual work. \textbf{SPICE} automates labeling at \textbf{19,000× lower monetary cost than human annotation}. Asynchronous Docker execution eliminates pipeline bubbles and delivers a \textbf{3× RL speedup}. Altogether, these innovations enable RepoForge-8B-Agent which is trained on Qwen3-8B to achieve \textbf{17.4\% on SWE-Bench-Verified}, setting a new state of the art for models under 8B parameters.
We also share key lessons learned about data quality, training strategies, and the unique challenges of scaling small models for software engineering tasks.

\begin{figure}[htbp]
\centering
\includegraphics[width=\linewidth]{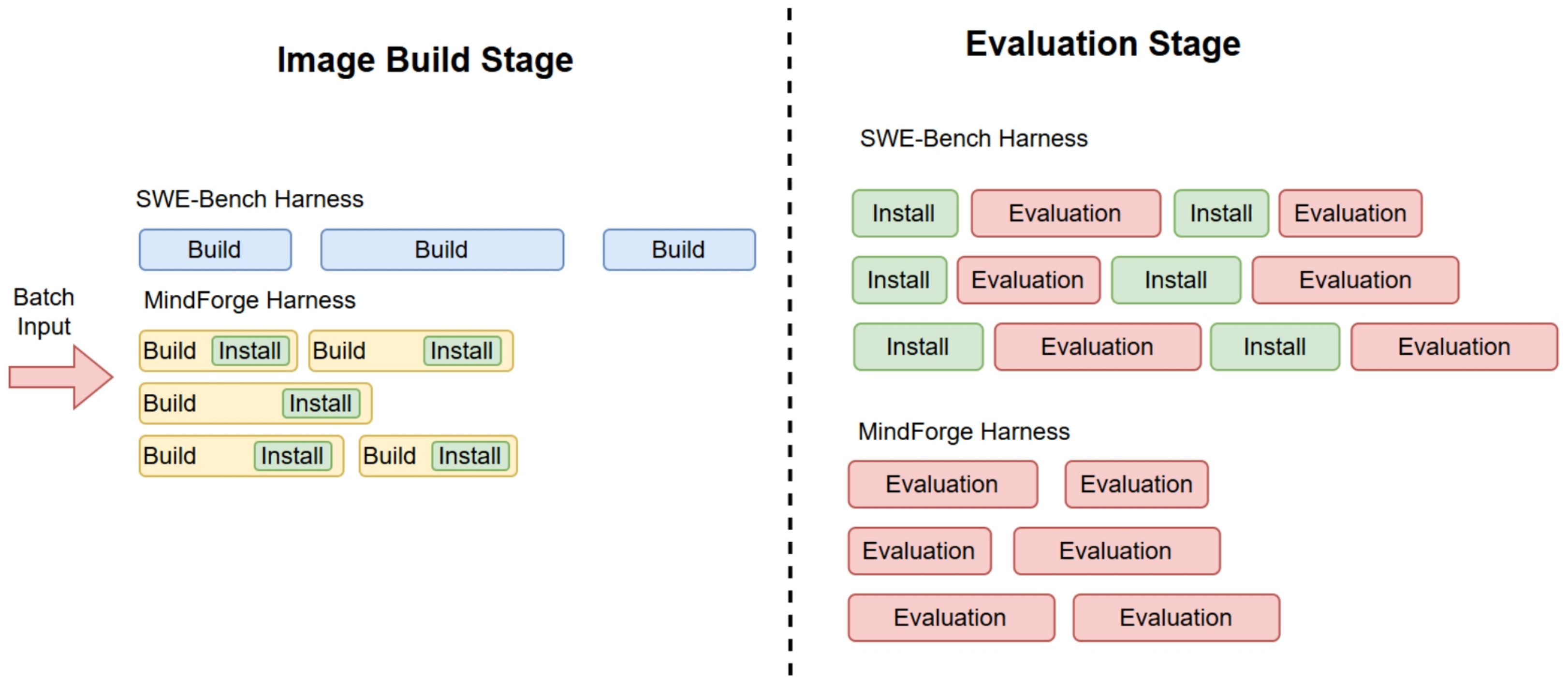}
\caption{This diagram illustrates the RepoForge pipeline, designed to address key challenges in training and evaluating reinforcement learning (RL) agents for software engineering tasks.}
\label{fig:RepoForge_pipeline}
\end{figure}  
\section{Challenges}
\label{sec:challenges}

Building production-scale SWE agents requires addressing challenges that extend far beyond just the evaluation harness, touching every aspect of the training pipeline. We identify five critical challenges that create fundamental barriers to scaling SWE agent training:

\subsection{Challenge 1: High Storage Costs of Container-Based Sandbox Evaluation}
\label{subsec:challenge1}

In many existing SWE evaluation setups, each task is isolated in its own Docker image for reproducibility and safety. However, a typical image averages about 1.4 GB. When scaled to hundreds or thousands of tasks, the required storage grows quickly and can overwhelm normal infrastructure.

Prior work highlights how costly containerized environments can be: for example, SkyRL~\citep{skyrl2024}reports that each running environment may consume over one CPU and around 7 GB of storage, while a similar RL framework, rLLM~\citep{rllm2025}, requires roughly 6 TB of disk space to store thousands of images. Even a moderate setup, such as 16 tasks with 8 rollouts each, can use over 100 CPUs and nearly 1 TB of disk space. Running just 500 SWE‑Bench‑Verified tasks already needs about 700 GB. Scaling to RepoForge’s 7,304 tasks would exceed 10 TB if the same approach were used.

The problem gets worse with reinforcement learning and rejection sampling, where many containers run in parallel. For example, 128 concurrent tasks at 7 GB each need around 896 GB just for runtime. Running all 7,304 tasks with both base and runtime images would require around 61 TB, far beyond typical hardware.

When storage becomes the bottleneck, CPUs sit idle waiting for disk resources, severely limiting training speed and efficiency.

\begin{figure}[htbp]
\centering
\includegraphics[width=0.8\linewidth]{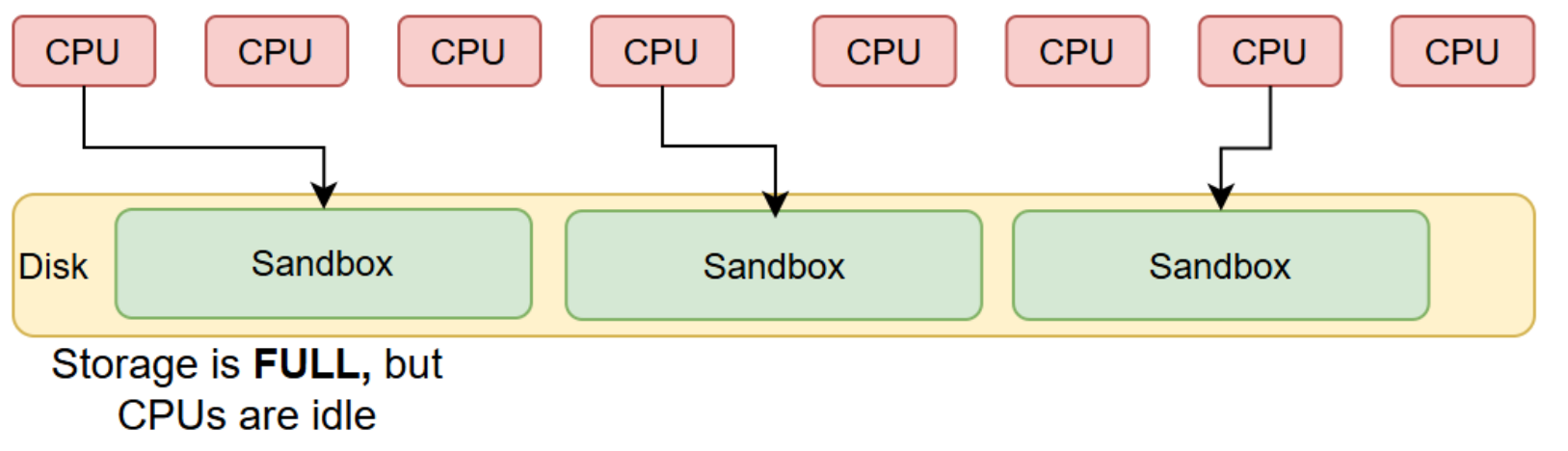}
\caption{In real training scenarios, the worker concurrency is highly constrained to disk utilization.}
\label{fig:disk_utilization}
\end{figure} 

\subsection{Challenge 2: Inefficient and Sequential Reward Evaluation Pipelines}
\label{subsec:challenge2}

Execution feedback is critical for training software engineering agents because, unlike typical NLP tasks scored with static metrics, their outputs must actually run in real environments. A patch is only correct if it compiles, passes all tests, and leaves existing functionality intact. This requirement for real execution creates unique infrastructure demands.

Existing systems like SWE-Bench~\citep{jimenez2024swebench} and SWE-Gym~\citep{swegym2024} handle this with sequential, blocking pipelines. Each task must fully finish building its Docker environment, then installing dependencies, and only then run evaluations. There is no overlap between these stages and no reuse of prior work. Managing dependencies becomes complex (1,452‑line configuration file), often involving massive configuration files and separate Docker images for each instance. Because every instance reinstalls its dependencies from scratch, resources are wasted as CPUs and memory remain idle during blocking I/O.

These problems worsen when scaling up for reinforcement learning. RL requires frequent reward calculations, but sequential evaluation causes each instance to take 2 to 10 minutes, so multi‑turn training with 20–50 iterations can stretch into days.

The engineering challenge is that repositories vary widely in languages and dependencies, each must run in an isolated environment, results must be reproducible, network calls can slow down installs, and large‑scale RL needs hundreds of evaluations per hour. Traditional frameworks, designed for simpler static benchmarks, cannot handle this dynamic and resource‑intensive execution at scale.

\subsection{Challenge 3: Limited Availability of High-Quality SWE Training Data}
\label{subsec:challenge3}

High‑quality, executable SWE training data is very limited. Most public benchmarks only offer a few hundred tasks that are static, outdated, and built for one‑time evaluation rather than large‑scale training. They depend heavily on manual environment setup and human labeling, which is slow, costly, and difficult to scale.

For teams training LLMs for software engineering, this scarcity causes serious problems. There aren’t enough varied tasks to teach models real‑world challenges, annotations are inconsistent and subjective, and it can take months of manual work just to collect a few thousand usable examples. This leads to stalled training pipelines, models that plateau, and high costs that block scaling.

Existing systems highlight this challenge. SWE-bench~\citep{jimenez2024swebench} uses manual Dockerfile setups for 500 tasks, requiring heavy labor and often leading to compatibility issues. SWE-gym~\citep{swegym2024} spent over 200 hours to build about 2,400 tasks, still fully manual and not scalable. Skywork-SWE~\citep{skywork2024swe} adds some automation with a three-tier Docker hierarchy, but still needs manual configuration templates. SWE-bench Live~\citep{swebench_live2024} relies on an LLM agent to set up environments, but it doesn't learn from previous attempts. SWE-factory~\citep{swefactory2024} uses multiple agents with memory pooling, but remains complex and still needs manual checks. These approaches show how difficult it has been to create large, reliable SWE datasets.

\subsection{Challenge 4: High Cost and Subjectivity of Manual Data Labeling}
\label{subsec:challenge4}

Labeling SWE training data is a major bottleneck because it demands expert judgment on problem clarity, test validity, and fix difficulty. OpenAI’s SWE-bench Verified~\citep{swebench_verified2024}  addressed this by manually filtering 500 high-quality examples from over 12,000 candidates. On this curated set, GPT-4o achieved a 33.2\% pass rate, more than doubling prior open-source results.

Each SWE-bench Verified task had to meet three strict criteria: the issue statement had to be unambiguous, tests had to cover all valid solutions without rejecting correct fixes, and an experienced engineer should be able to implement the patch in under an hour. Reaching these standards required months of human review and cost thousands of dollars per thousand labels.

Unlike simpler NLP annotations, SWE labeling involves deep repository exploration—tracing dependencies across multiple files, interpreting build and test pipelines, and actually running patches to confirm correctness. Reinforcement-learning pipelines demand tens of thousands of such labeled instances, making manual annotation economically and practically impossible. Continuous changes in real-world codebases only add to the challenge, as labels can become outdated with dependency updates, requiring ongoing re-labeling.

\subsection{Challenge 5: Agentic Multi-Turn Reinforcement Learning Pipeline Bottlenecks}
\label{subsec:challenge5}

Multi‑turn reinforcement learning at the repository level relies on three stages: the model generates actions, interacts with an environment, and receives rewards. Scaling this to 20–50 iterations per trajectory, as required by the OpenHands scaffold, makes it difficult to keep the system efficient, stable, and well‑utilized.

\textbf{One major issue is resource management}. Frameworks like SkyRL-OpenHands~\citep{skyrl2024} use a two-stage Docker build that adds an internal server layer, increasing each image by 4–10 GB and quickly consuming disk space. Others, like R2EGym~\citep{r2egym2024}, lack distributed training support, so every machine must store full images, adding even more storage pressure.

\textbf{Another issue is pipeline bubbles}. Multi‑container workflows over many iterations create overhead and idle time. Model generation and environment execution rely on different resources, and if there are too few rollout requests or too few concurrent executions, CPUs sit idle. Latency from network calls, image builds, and I/O adds to the delay. Worse, the pipeline often waits for all trajectories in a batch to finish; if one task takes much longer than the others, it creates a long‑tail delay that stalls every worker and reduces overall throughput.
\section{Our Method}
\label{sec:method}

To address the challenges outlined in Section~\ref{sec:challenges}, we present RepoForge, a comprehensive solution that tackles each bottleneck through innovative technical approaches. Our method consists of five key components that work together to create an efficient, scalable pipeline for training SWE agents.

\subsection{Solution to Challenge 1: Storage-Efficient Image Pruning \& Minimal Runtime Environments}
\label{subsec:solution1}

To address the storage bottleneck, we leverage \textbf{Image Dependency Pruning}, an automatic algorithm to merge multiple images into one single image, and \textbf{Minimal Runtime Environments}, stripping out any unused dependencies so we can reuse these images for multi-instances.

\subsubsection{Image Dependency Pruning}
To fix this, RepoForge uses an automatic process called Image Dependency Pruning. This process looks at all the Docker images and finds shared dependencies between them. When two images share enough in common and all tests still pass, one image can be rebased to reuse the other instead of keeping its own copy. A custom algorithm goes through all instances and keeps trying to merge them with earlier or later images until no more merges are possible. With this method, thousands of tasks that would normally need thousands of heavy images can instead share a much smaller number of optimized images as illustrated in Figure \ref{fig:image_reuse}.

\begin{figure}[htbp]
\centering
\includegraphics[width=\linewidth]{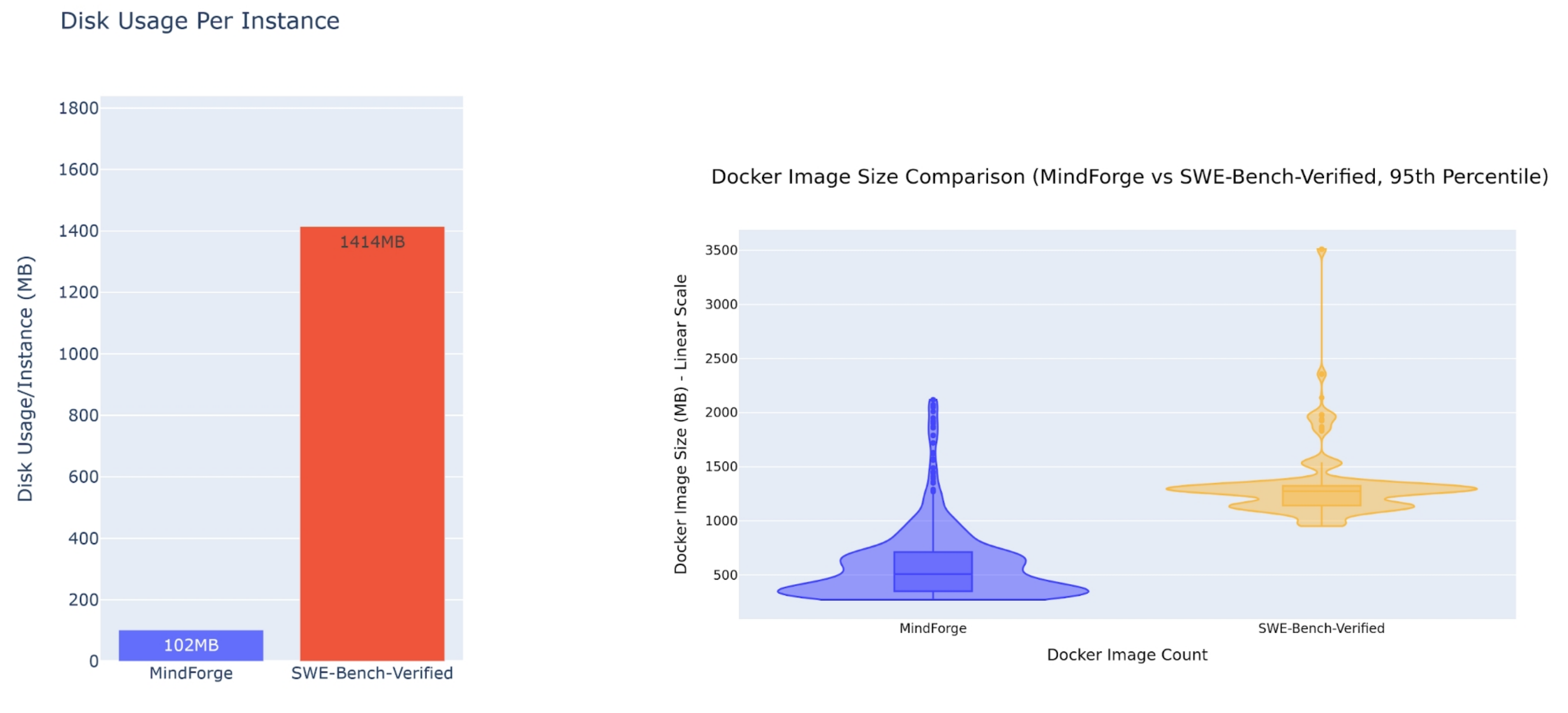}
\caption{Disk usage per instance in the RepoForge framework versus SWE-Bench-Verified. Left: Per Instance, Right: Single Image size distribution}
\label{fig:disk_usage_comparison}
\end{figure} 
\subsubsection{Minimal Runtime Environments}
On top of that, RepoForge builds Minimal Runtime Environments. Instead of using large base images like full Ubuntu builds, it starts with very slim images and only installs exactly what is needed for each task. Any unnecessary packages, compilers, or tools are removed. This cuts down the size of each image dramatically.
\begin{figure}[htbp]
\centering
\includegraphics[width=0.8\linewidth]{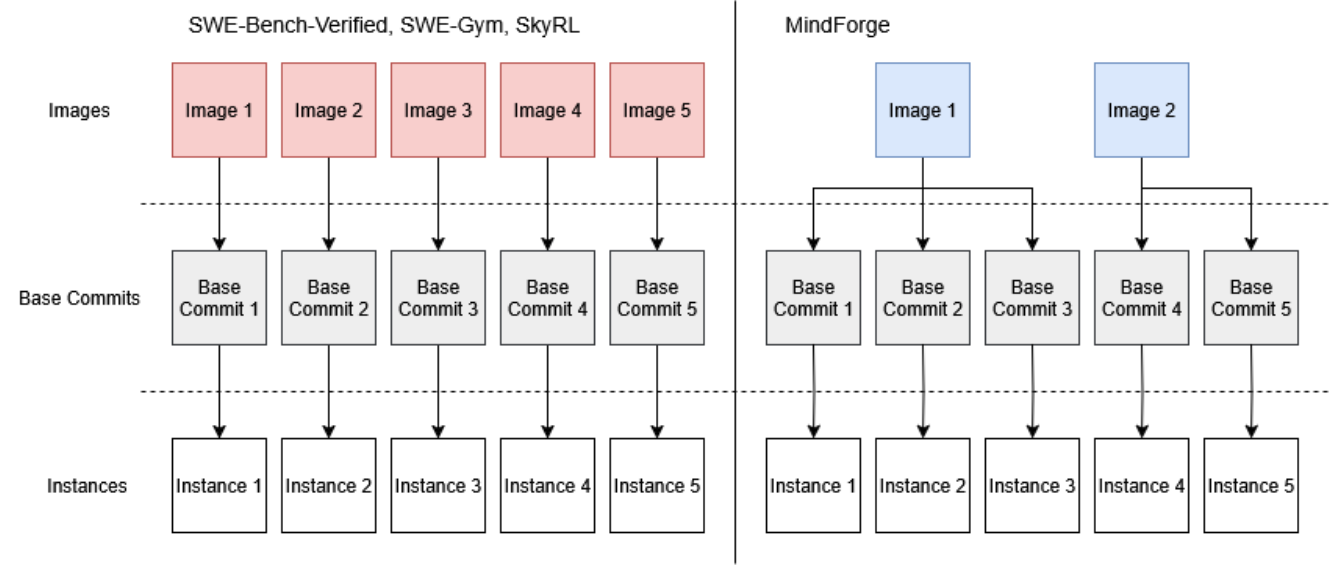}
\caption{In RepoForge Framework, multiple base commits can work on the same image, greatly reducing the disk requirement.}
\label{fig:image_reuse}
\end{figure} 

With these two methods combined, the average image size per instance drops from \textbf{1.4 GB to just 102 MB}. In total, \textbf{7,304 tasks only require 937 optimized images}, with each image reused across multiple tasks. This means less disk space is used, CPU resources are no longer idle waiting on storage, and the whole training pipeline runs much faster and more efficiently.

\subsection{Solution to Challenge 2: RepoForge Harness - A Ray-Powered Distributed Efficient Evaluation Engine}
\label{subsec:solution2}

To improve the efficiency of evaluating rewards from code execution feedback, we built \textbf{RepoForge Harness}, inspired by the SWE-Bench evaluation harness~\citep{jimenez2024swebench}. RepoForge Harness is fully distributed (based on Ray) and supports streaming, overlapping stages while reusing build artifacts across task instances, achieving \textbf{10$\times$ build speedup and 3$\times$ evaluation speedup}.

\begin{figure}[htbp]
\centering
\includegraphics[width=\linewidth]{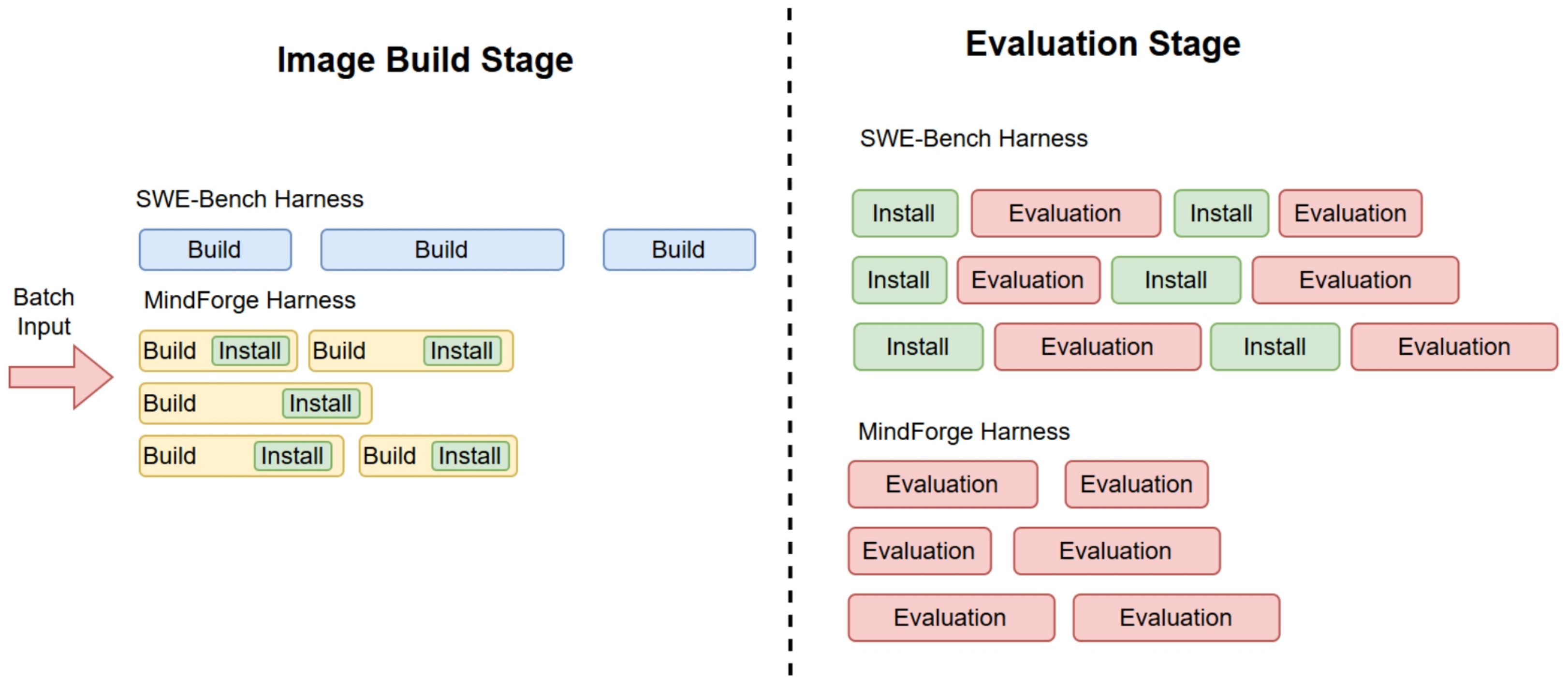}
\caption{In the default SWE-Bench harness, container images are built one at a time, in a strictly sequential manner. During evaluation, RepoForge pre-installs all required runtime dependencies for each unique task instance so the evaluation can run immediately.}
\label{fig:build_pipeline}
\end{figure}

One key improvement is that RepoForge Harness enables a parallel, streaming style build pipeline. Instead of blocking on a single image, RepoForge builds many images at the same time using aiodocker as illustrated in Figure \ref{fig:build_pipeline}. In addition, dependencies are installed during the build stage, so evaluation starts immediately without extra network calls or repeated installs. This is very efficient especially when the training environment is behind corporate network wall.

\begin{figure}[htbp]
\centering
\includegraphics[width=0.8\linewidth]{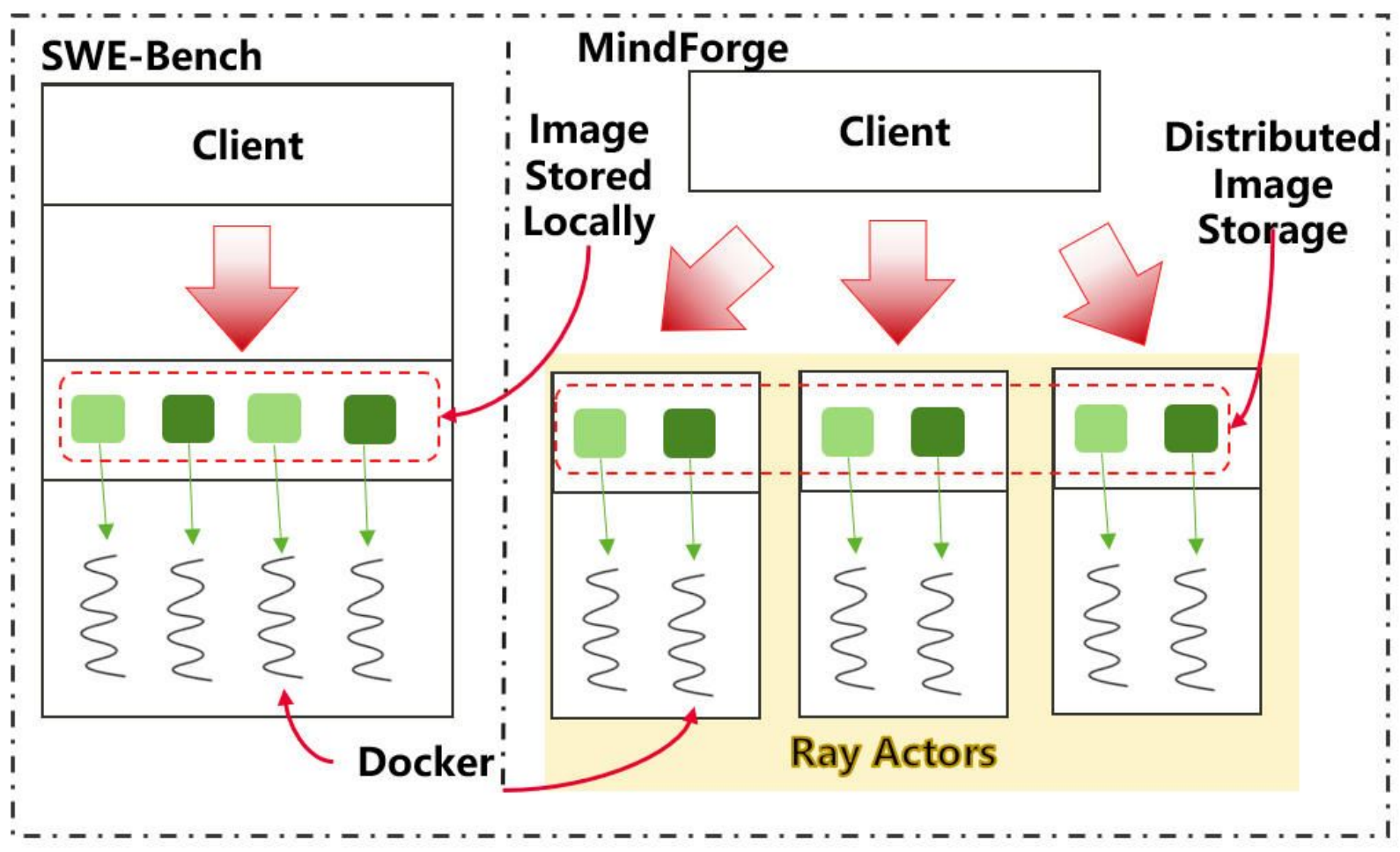}
\caption{RepoForge leverages Ray Actors to scale up the sandbox server across multiple nodes.}
\label{fig:ray_actors}
\end{figure}

Another improvement is Ray‑powered~\citep{ray2018} distributed execution. Instead of handling everything on one local machine, RepoForge uses Ray actors to start sandboxes on multiple machines at once. Each actor can reuse shared container images, so the system avoids redundant builds and handles task instances in parallel as illustrated in Figure \ref{fig:ray_actors}.

In tests, over 96 percent of evaluations finish within two minutes, with a hard cap of 120 seconds. Average evaluation time drops from 2–10 minutes in the old system to about 75 seconds. The system can scale up to 64 workers, supports automatic pooling and intelligent caching, and works across multiple programming languages like Python and C++ without special setup.

\begin{figure}[htbp]
\centering
\includegraphics[width=0.8\linewidth]{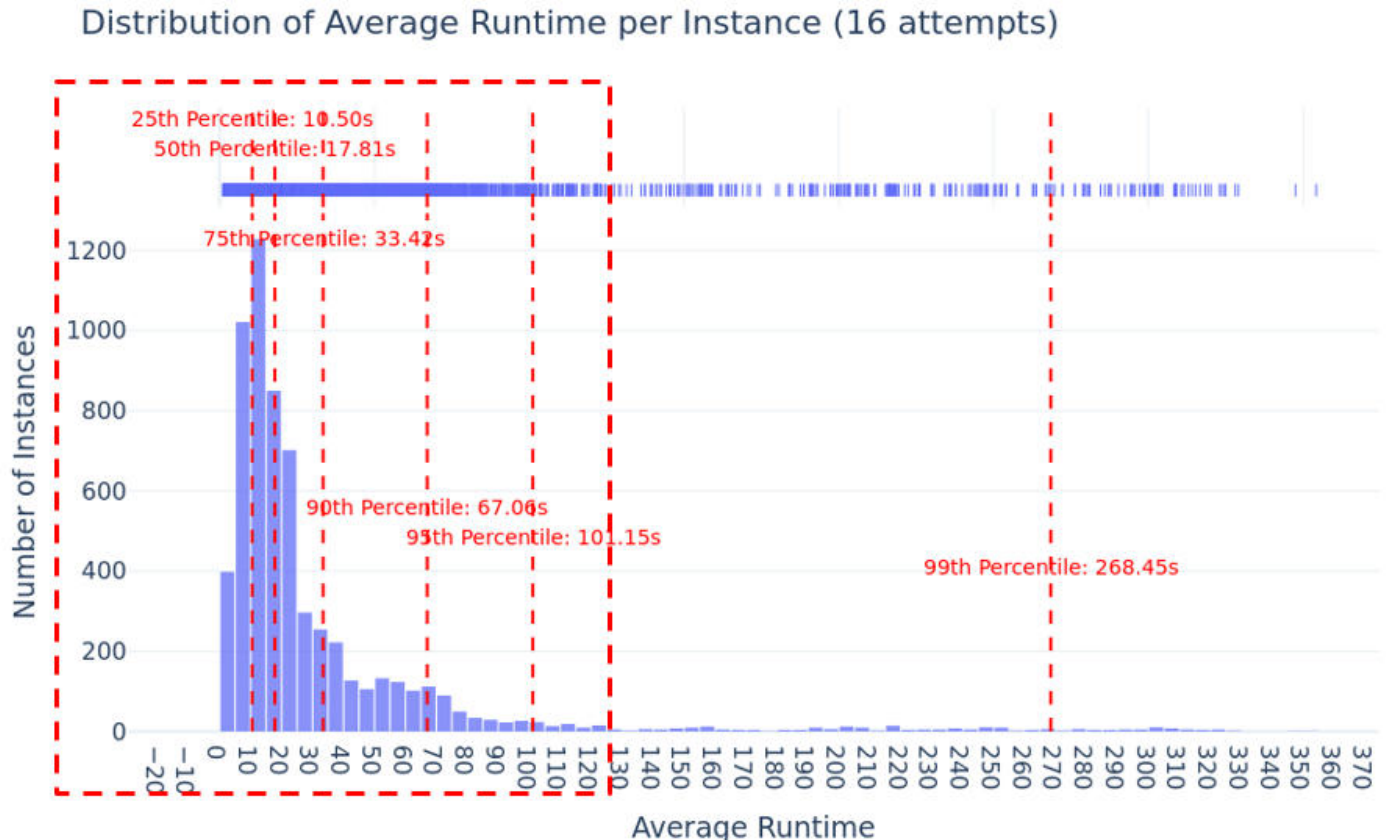}
\caption{Averaged over 16 runs, RepoForge instance evaluations remain consistently fast, over 96\% complete within two minutes.}
\label{fig:evaluation_times}
\end{figure} 

\begin{table}[htbp]
\centering
\caption{Comparison of SWE-Bench \& SWE-Gym Harness system vs RepoForge Harness}
\label{tab:harness_comparison}
\begin{tabular}{@{}lcc@{}}
\toprule
\textbf{Feature} & \textbf{SWE-Bench \& SWE-Gym} & \textbf{RepoForge} \\
\midrule
Architecture & Sequential & Distributed (Ray) \\
Concurrency & Single-threaded & Up to 64 workers \\
Image Reuse & None & Intelligent caching \\
Container Management & Manual & Automatic pooling \\
Install Timing & After build & During build \\
Execution Model & Blocking & Fully async \\
Image Build Time (Mean) & Average 537s & Average 58s \\
Evaluation time (Mean) & Average 56s & Average 17s \\
Multi-Language Support & No & Yes \\
\bottomrule
\end{tabular}
\end{table} 

\begin{figure}[htbp]
\centering
\includegraphics[width=\linewidth]{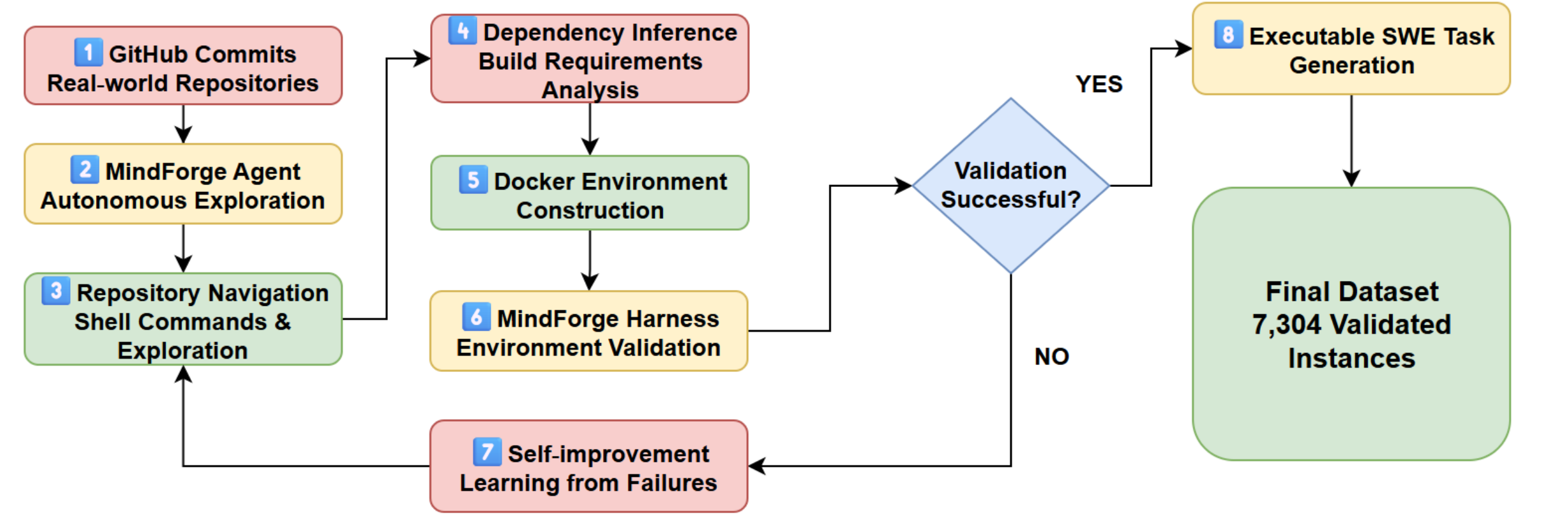}
\caption{Overview of the RepoForge automated pipeline for dependency inference, Docker environment construction, validation, self-improvement, and SWE task generation.}
\label{fig:automated_pipeline}
\end{figure} 

\subsection{Solution to Challenge 3: RepoForge Foundry - Fully Autonomous Data Generation at Scale}
\label{subsec:solution3}

To break through the data wall, we built \textbf{RepoForge Foundry} — a fully autonomous system that creates evaluation-ready SWE environments with minimal human intervention.

Instead of relying on a small, hand‑labeled benchmark, RepoForge Foundry actively explores real‑world GitHub commits, reconstructs their build and test environments, and validates them into rich, executable tasks — all without human intervention.

RepoForge Foundry leverages a multi‑agent ReAct framework to autonomously ingest real‑world GitHub repositories and issues, infer their full build graph, and generate optimized, isolated Docker environments. Then, RepoForge Harness (described in the solution of Challenge 2) validates each image with end‑to‑end FAIL\_TO\_PASS and PASS\_TO\_PASS tests and automatically rolls back and learns from any failures. For a detailed workflow, see Figure \ref{fig:automated_pipeline}.

We end up with an automated SWE environment construction pipeline that supports multiple programming languages and hybrid build systems, guarantees test correctness by actually executing tests. By reducing manual efforts, RepoForge Foundry enables SWE data as a renewable fuel for training.

The result is \textbf{7,304 fully validated and annotated SWE task instances} generated automatically, powered by only \textbf{937 optimized Docker images} with built-in reuse. Furthermore, we are able to label rich information on these instances at a cost much lower than manual curation.

\begin{table}[htbp]
\centering
\caption{Comparison of existing SWE environment-generation systems vs RepoForge Agent}
\label{tab:systems_comparison}
\footnotesize
\begin{tabular}{@{}lcccccc@{}}
\toprule
\textbf{System} & \textbf{Manual Work} & \textbf{Scale} & \textbf{Docker Opt.} & \textbf{Validation} & \textbf{Learning} & \textbf{Eval-ready*} \\
\midrule
SWE-bench & 100\% Manual & 500 instances & None & Human & None & Yes \\
SWE-gym & 200+ hrs manual & 2.4K instances & None & Human & None & Yes \\
Skywork-SWE & Template creation & 10K instances & Basic hierarchy & Semi-auto & None & Yes \\
SWE-Builder & Manual verification & Unknown & None & Mixed & Memory pool & Yes \\
Repo2Run & 0\% Manual & 420 instances & None & Fully automated & None & No \\
\textbf{RepoForge Agent} & \textbf{0\% Manual} & \textbf{7.3K instances} & \textbf{937 optimized} & \textbf{Fully automated} & \textbf{Self-improving} & \textbf{Yes} \\
\bottomrule
\end{tabular}
\vspace{0.2cm}
\begin{flushleft}
\textbf{*Evaluation-ready} only if it correctly runs the designated FAIL\_TO\_PASS and PASS\_TO\_PASS tests both before and after applying the golden patch, ensuring meaningful evaluation.
\end{flushleft}
\end{table} 

\subsection{Solution to Challenge 4: SPICE - Automated Difficulty Assessment at Scale}
\label{subsec:solution4}

Our SPICE (Structured Problem Instance Classification Engine)~\citep{spice2024} eliminates the manual labeling bottleneck by automatically evaluating task difficulty and quality across four dimensions: code complexity, repository structure, test coverage, and solution patterns. It then assigns objective difficulty scores based on quantitative metrics, filters for high-value instances, and continuously refines its assessments using model performance data at a tiny fraction of the cost of human annotation.

Key innovations in SPICE include its reliance on concrete code analysis instead of subjective judgment, a fully automated difficulty‑scoring pipeline, and a scalable architecture that applies consistent criteria across thousands of instances. This approach delivers uniform, reliable labeling at scale and reduces annotation costs significantly compared to traditional manual methods.

With SPICE, we automatically labeled and difficulty-assessed over \textbf{7,000 instances}, achieved a \textbf{19,600$\times$ cost reduction} compared to manual labeling while maintaining \textbf{87.3\% accuracy} on issue clarity and \textbf{68.5\% on test coverage}, and enabled rapid iteration for new datasets.

\subsection{Solution to Challenge 5: RepoForge-OpenHands Scaffold}
\label{subsec:solution5}

To address challenge 5, we introduced the \textbf{RepoForge-OpenHands scaffold}, which inherits most of the original OpenHands scaffold features and implemented four additional engineering optimizations. As a result, we achieved a \textbf{3$\times$ speedup}, making it more suitable for long-horizon agentic RL training.

\subsubsection{Direct Docker Exec Integration}
RepoForge-OpenHands mounts its OpenHands-styled Python tools into containers and the runtime invokes them via docker exec, removing the embedded server and second build stage (4 - 10 GB per image and additional build time) by back-porting OpenHands utilities to Python 3.5, yielding a \textbf{$\sim$5$\times$} speedup and \textbf{cutting off 80\% of disk storage} during the training.

\subsubsection{Remote Sandbox Server}
We decouple the execution of the environment from the training with a Ray-managed remote sandbox that runs two concurrent pipelines of up to 32 agents each, \textbf{$\sim$80–100 isolated containers per host}, allowing for scalable parallel isolation without impacting the training process.

\subsubsection{Fully Asynchronous I/O}
By replacing Docker’s blocking HTTP client calls with non‑blocking libraries (aiodocker, aiohttp), we eliminate GIL‑induced serialization, achieve true parallelism in environment dispatch, and significantly reduce I/O overhead under high concurrency.

\subsubsection{Dynamic Producer–Consumer Rollout}
Using Ray's worker queue model, we created a producer-consumer model, where workers dynamically poll (instance, trajectory) pairs instead of fixed subsets, eradicating long-tailed stalls in balanced batching and delivering a \textbf{1.4$\times$ speedup} on representative test batches.

After applying the above optimization, the results in Figure 15 show a 3× end‑to‑end speedup compared with the vanilla OpenHands, a \textbf{5× faster tool invocation} with over 8\textbf{0\% storage reduction}, and a \textbf{1.4× throughput gain} by smoothing idle time and improving resource use in long-tail batches.

\begin{figure}[htbp]
\centering
\includegraphics[width=0.6\linewidth]{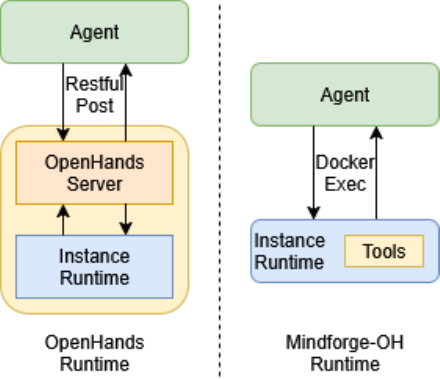}
\caption{Comparison of the OpenHands server-based runtime versus the Docker-exec-based RepoForge-OpenHands runtime architectures.}
\label{fig:runtime_architectures}
\end{figure}

\begin{figure}[htbp]
\centering
\includegraphics[width=\linewidth]{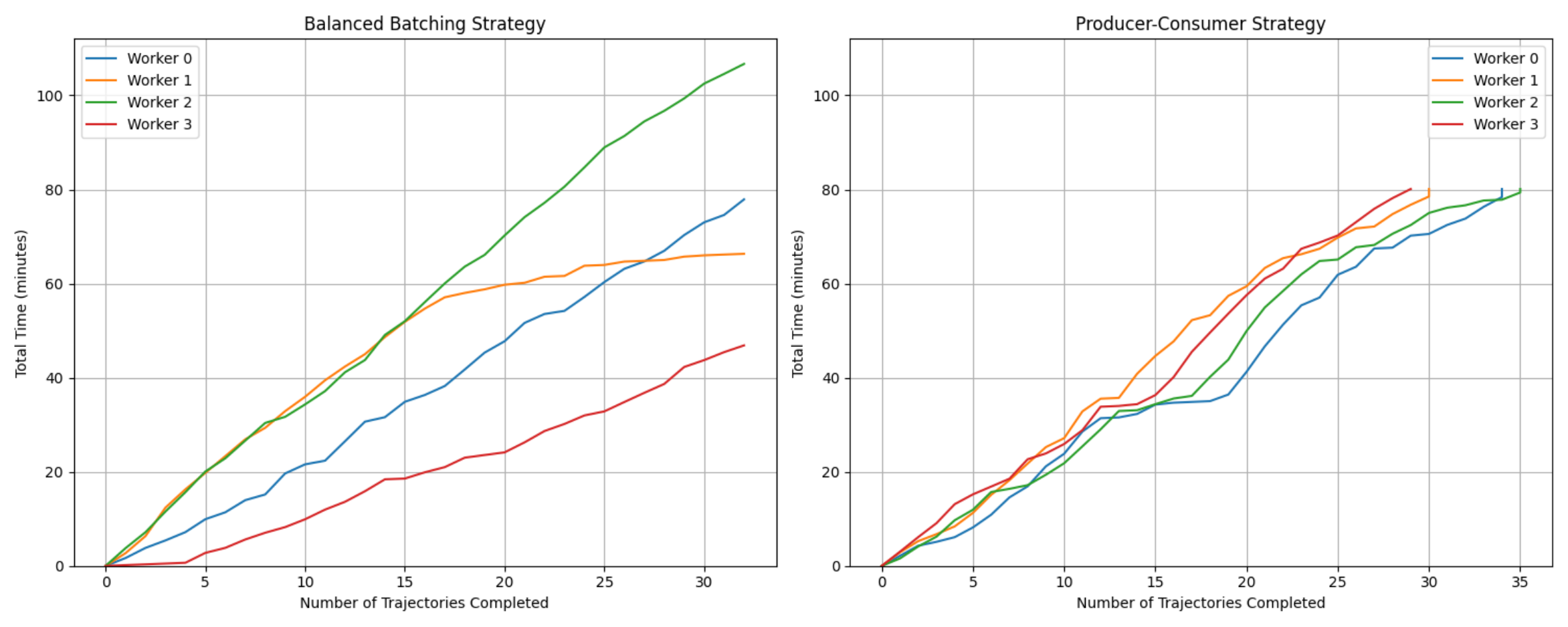}
\caption{Comparison of rollout distribution strategies - Balanced Batching vs. Producer-consumer on RepoForge-OpenHands scaffold.}
\label{fig:rollout_strategies}
\end{figure} 

\begin{figure}[htbp]
\centering
\includegraphics[width=\linewidth]{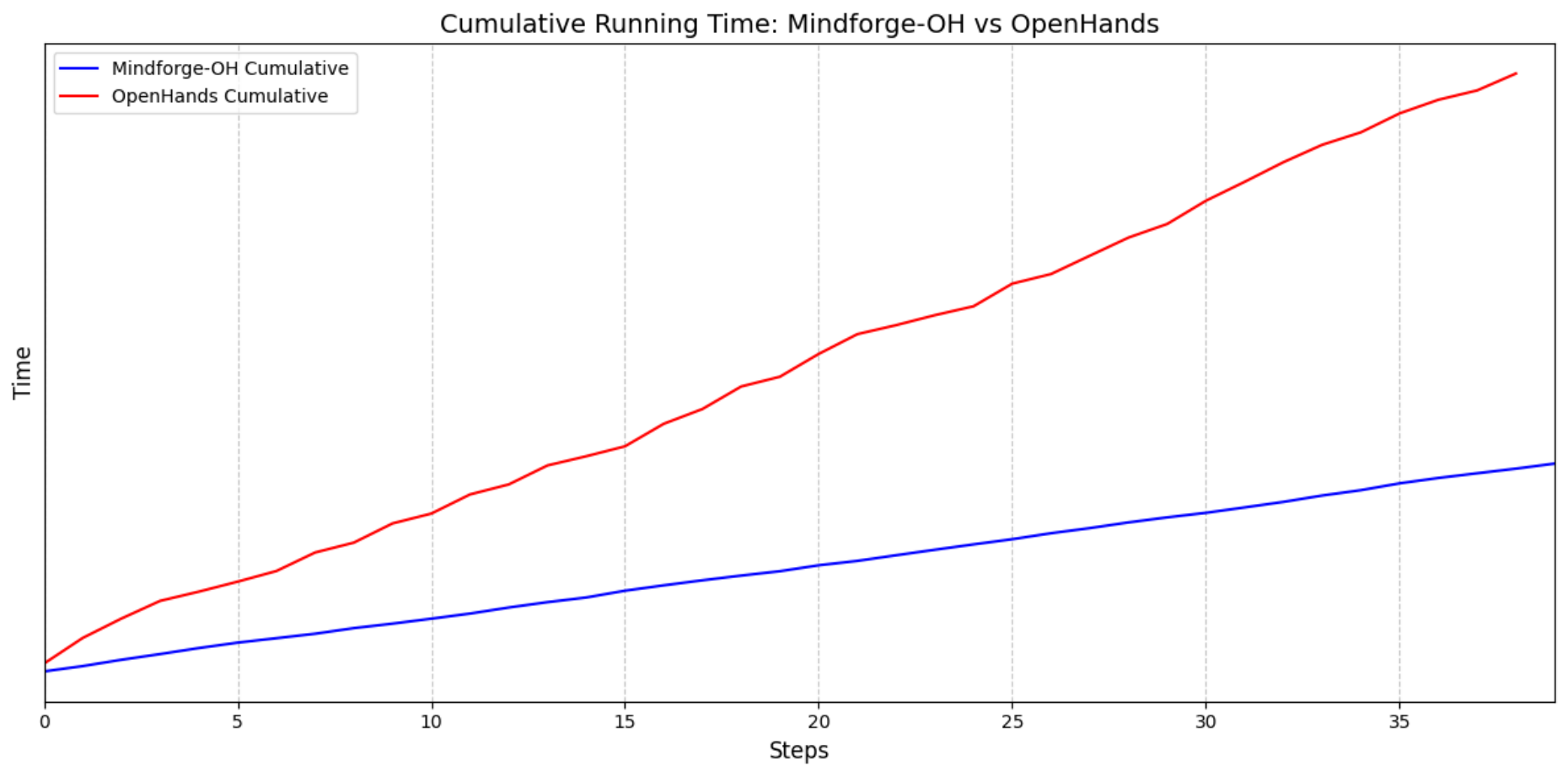}
\caption{Comparison of running times - OpenHands Scaffold vs. RepoForge-OpenHands Scaffold on 40 training steps with the same dataset (not including build).}
\label{fig:running_times}
\end{figure}  
\section{Training Recipe}
\label{sec:training}

With over 7,304 executable environments, the next challenge was ensuring data quality. Quantity was no longer a problem, but we needed tasks that were well-specified, properly tested, and of manageable difficulty. Our training approach consists of two main phases: Supervised Fine-Tuning (SFT) followed by Reinforcement Learning (RL).

\subsection{Supervised Fine-Tuning (SFT)}
\label{subsec:sft}

First, we used SPICE to score each raw instance on clarity, test validity, and difficulty. Only those scoring 1 or lower on all three metrics were kept, removing ambiguous, poorly tested, or overly complex tasks. This produced a pool of clean, reliable tasks.

Next, we ran rejection sampling with OpenHands-32B-Agent~\citep{openhands_lm_32b2024} under the RepoForge Harness. For each filtered instance, the agent generated up to eight patches. Each patch was applied and immediately tested. Only patches that met both PASS\_TO\_PASS and FAIL\_TO\_PASS criteria were kept. To avoid overfitting and hallucination, every run used a randomized working directory and an isolated container.

This process produced 1,202 high-quality instances, each with a complete multi-turn trajectory, SPICE metadata, and validated test suites. These were used to fine-tune RepoForge-8B-Agent over one epoch, creating a strong starting point for reinforcement learning.

\subsection{Reinforcement Learning (RL)}
\label{subsec:rl}

For RL, we built on this SFT model using the RepoForge-OpenHands scaffold~\citep{openhands2024} with CodeAct and AsyncRollout. The model used three core tools: \texttt{execute\_bash} to run shell commands, \texttt{str\_replace\_editor} to edit files, and \texttt{finish} to end a trajectory. Rewards were simple and binary—1 if all tests passed, 0 otherwise.

Training used a curated dataset of 160 instances that the teacher model had already solved during SFT. This ensured the model started with tasks it understood, allowing RL to focus on refining behavior rather than learning from scratch. We trained for 40 steps over two epochs with a temperature of 0.5, a KL-divergence coefficient of 0.001, and a cap of 35 iterations per trajectory.

The combination of high-quality SFT data and carefully curated RL tasks enabled our model to achieve significant performance improvements while maintaining stability throughout the training process.

\begin{figure}[htbp]
\centering
\includegraphics[width=0.8\linewidth]{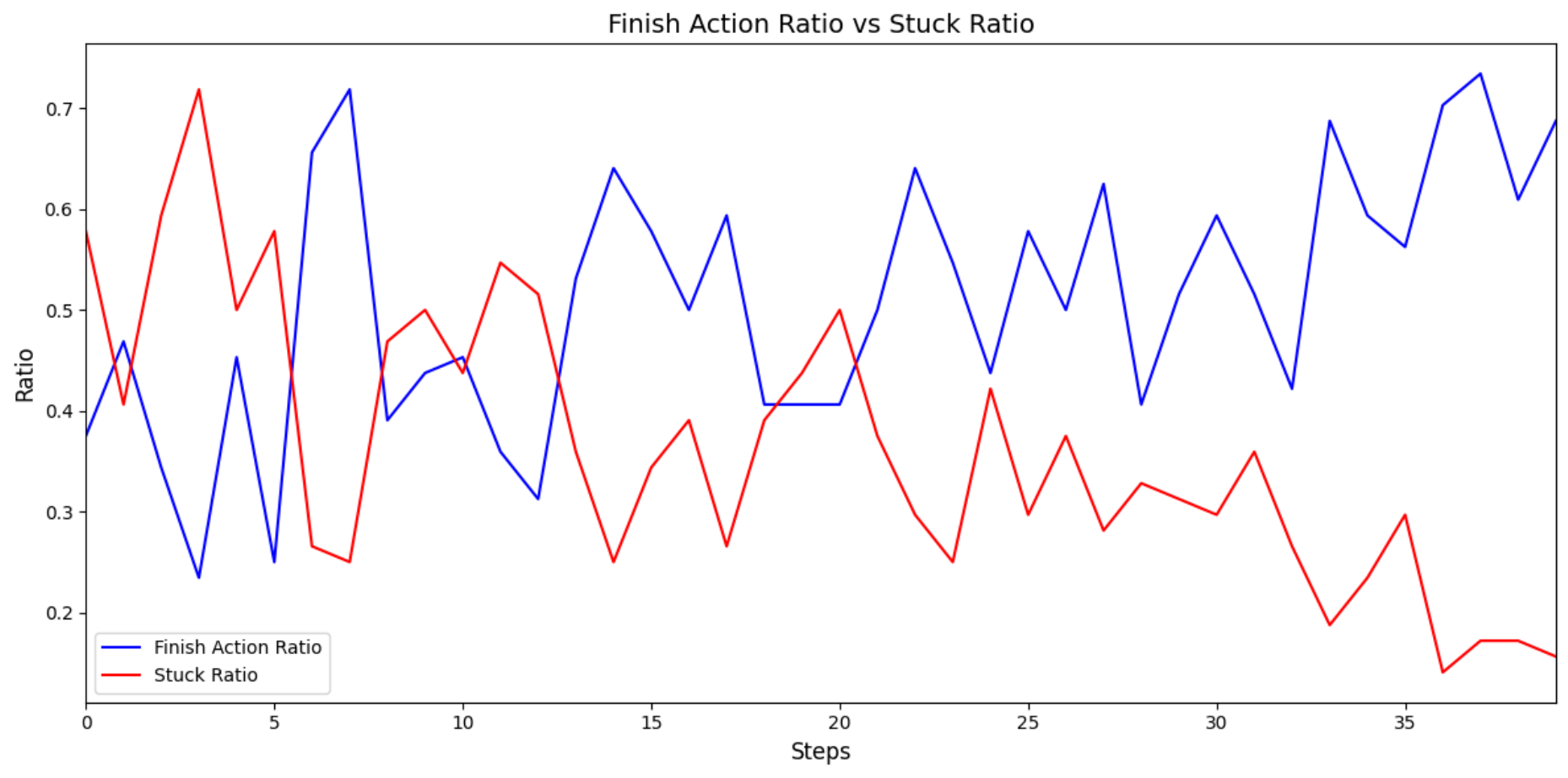}
\caption{Action finish ratio and stuck ratio during the training.}
\label{fig:training_ratios}
\end{figure}  
\section{Evaluation}
\label{sec:experiments}

We evaluate \textbf{RepoForge-8B‑Agent} on both the\textbf{ RepoForge‑OpenHands} and the \textbf{official OpenHands framework}\textbf{}, summarizing resolve rates on SWE‑Bench‑Verified\textbf{} instances across various models and configurations. A successfully resolved instance means the Agent model generated and applied a code patch, with all the unit tests corresponding to the problem statement having been run and passed. In our table, the “RepoForge-OpenHands” percentages report resolve rate performance under our RepoForge harness, “OpenHands” percentages reflect results from the official OpenHands harness main branch @ <commit-id> or release version xyz, and “reported” percentages reproduce scores as published in the original papers. The “thinking” designation denotes whether a model employs explicit reasoning (e.g., chain‑of‑thought or scratchpad) or runs without such reasoning, which means the model does not natively support outputting thinking content or it has been disabled via the chat template.

\subsection{Main Results}
\label{subsec:main_results}

\textbf{RepoForge-8B-Agent (SFT+RL)} achieves \textbf{17.4\% on RepoForge-OpenHands} and \textbf{16.4\% on OpenHands}, setting a new state-of-the-art among $\leq$8B models, even though it does not use reasoning steps. This represents a significant achievement in the field of software engineering agents, demonstrating that smaller models can achieve competitive performance when trained with the right pipeline and data.

\subsection{Ablation Studies}
\label{subsec:ablation}

RL adds significant performance gains: SFT-only training plateaued at 12.7\% on RepoForge-OpenHands, while combining SFT with RL pushed accuracy to 17.4\%. This 4.7 percentage point improvement highlights the importance of the reinforcement learning phase in our training pipeline.

RepoForge-8B-Agent matches or surpasses several 14B–32B-scale models, proving that small models can punch well above their weight when properly trained and optimized.

\subsection{Performance Analysis}
\label{subsec:performance}

Our evaluation demonstrates several key findings:

\begin{itemize}
\item \textbf{Model Size vs. Performance}: Despite being only 8B parameters, RepoForge-8B-Agent achieves competitive performance with much larger models, highlighting the effectiveness of our training approach.

\item \textbf{Infrastructure Impact}: The RepoForge Harness consistently shows improved performance compared to standard evaluation frameworks, suggesting that evaluation infrastructure itself can impact measured performance.

\item \textbf{Training Strategy Effectiveness}: The combination of high-quality SFT data followed by targeted RL training proves more effective than either approach alone.
\end{itemize}

These results validate our hypothesis that addressing infrastructure bottlenecks and improving data quality can enable smaller models to achieve state-of-the-art performance on challenging software engineering tasks.

\begin{table}[htbp]
\centering
\caption{Performance comparison of RepoMind-8B-Agent against prior coding agents}
\label{tab:performance_comparison}
\footnotesize
\begin{tabular}{@{}lcccc@{}}
\toprule
\textbf{Model} & \textbf{RepoForge\_OpenHands\%} & \textbf{OpenHands\%} & \textbf{Reported\%} & \textbf{Thinking} \\
\midrule
\textbf{RepoMind-Agent-8B(SFT+RL)} & \textbf{17.4} & \textbf{16.4} & \textbf{---} & \textbf{No} \\
RepoMind-Agent-8B(SFT Only) & 12.7 & 10.5 & --- & No \\
R2EGym-7B-Agent & 9.6 & 10.2 & 19 & No \\
Qwen3-8b(Non-thinking) & 5.0 & 4.2 & --- & No \\
Seed-Coder-8B-Instruct & 9.2 & 4.6 & 11.2 & No \\
SWE-Gym-OpenHands-7B-Agent & 10.8 & 11.4 & 14.6 & No \\
all-hands/openhands-lm-7b-v0.1 & 11.2 & 15.4 & 10.6 & No \\
\midrule
Qwen2.5-Coder-14B-Instruct & 4.8 & 3.2 & --- & No \\
Qwen3-32B(non-thinking) & 17.4 & 9.0 & --- & No \\
Qwen3-14B(non-thinking) & 8.0 & 3.0 & --- & No \\
deepseek-ai/deepseek-coder-33b-instruct & 1.6 & 1.6 & --- & No \\
\midrule
DeepCoder-14B-Preview & 8.0 & 6.6 & --- & Yes \\
Qwen3-8B & 17.8 & 11.4 & --- & Yes \\
Qwen3-32B & --- & 23.0 & --- & Yes \\
QwQ-32B & --- & 18.8 & --- & Yes \\
\bottomrule
\end{tabular}
\end{table}  
\section{Lessons Learned}
\label{sec:lessons}

From our experiments with RepoForge-8B-Agent, we learned several key lessons that shaped our design and provide valuable insights for future research in SWE agent training.

\subsection{Data Quality Matters Far More Than Quantity}
\label{subsec:lesson1}

Randomly sampled environments gave only 3.0\% accuracy after SFT, while SPICE~\citep{spice2024}-filtered high-quality data reached 12.7\%, a 4.2$\times$ improvement. In RL, random data gave 6.5\% while filtered data reached 7.8\%. This demonstrates that automated filtering techniques like SPICE~\citep{spice2024} and rejection sampling greatly improve both SFT and RL results.

The lesson here is clear: investing in data quality assessment and filtering mechanisms pays significant dividends in model performance. Rather than simply scaling up the quantity of training data, teams should focus on developing robust methods to identify and curate high-quality examples.

\subsection{Supervised Fine-Tuning Warm Start is Essential for RL}
\label{subsec:lesson2}

Running RL directly on the base model only reached 7.8\% accuracy, but adding a brief SFT stage boosted it to 17.4\%. This represents more than a 2$\times$ improvement and highlights the critical importance of the SFT phase.

SFT allows the model to succeed on some tasks early, providing useful rewards for RL to build on. Without this foundation, RL struggles to find positive signals in the sparse reward environment typical of software engineering tasks.

\subsection{Small Models Overfit Easily}
\label{subsec:lesson3}

Using too much SFT data hurt performance, with accuracy peaking at 12.7\% on 1,000 trajectories but dropping to 10.7\% on 2,000. This finding is particularly important for teams working with smaller models, as it suggests that careful dataset sizing and regularization are crucial to avoid overfitting.

For smaller models like our 8B parameter RepoForge-8B-Agent, finding the right balance between sufficient training data and avoiding overfitting requires careful experimentation and monitoring of validation performance throughout training.

\subsection{Infrastructure Optimizations Have Direct Impact on Model Performance}
\label{subsec:lesson4}

Our infrastructure improvements didn't just make training faster—they also improved model performance. The RepoForge Harness consistently showed better results than standard evaluation frameworks, suggesting that evaluation infrastructure itself can impact measured performance.

This insight highlights the importance of considering the entire training and evaluation pipeline when optimizing model performance, not just the model architecture and training algorithms.

\subsection{Implications for Future Work}
\label{subsec:implications}

These lessons suggest several directions for future research:

\begin{itemize}
\item \textbf{Automated Data Quality Assessment}: Developing more sophisticated methods for automatically assessing and filtering training data quality.
\item \textbf{Curriculum Learning}: Exploring how to optimally sequence training data from simple to complex tasks.
\item \textbf{Regularization Techniques}: Investigating regularization methods specifically designed for small models in software engineering domains.
\item \textbf{Infrastructure Co-design}: Considering evaluation and training infrastructure as part of the model development process rather than separate concerns.
\end{itemize} 
\section{Conclusions and Looking Ahead}
\label{sec:conclusion}

By unifying storage-efficient sandboxing, a Ray-powered~\citep{ray2018} evaluation harness, automated data generation, SPICE~\citep{spice2024}-based labeling, and a bubble-free RL scaffold, we have shown that even $\leq$8B models can reach new state-of-the-art performance on demanding benchmarks like SWE-Bench-Verified~\citep{swebench_verified2024}.

\subsection{Summary of Contributions}

Our work makes several key contributions to the field of software engineering agent training:

\begin{enumerate}
\item \textbf{Comprehensive Infrastructure Solution}: We addressed five critical challenges in SWE agent training through integrated technical innovations, demonstrating that infrastructure improvements can have significant impact on model performance.

\item \textbf{State-of-the-Art Results with Smaller Models}: RepoForge-8B-Agent achieves 17.4\% on SWE-Bench-Verified~\citep{swebench_verified2024}, setting a new state-of-the-art for models $\leq$8B parameters and proving that smaller models can be highly effective when properly trained.

\item \textbf{Scalable Data Generation Pipeline}: Our automated pipeline generated 7,304 executable environments from real GitHub commits with minimal human intervention, demonstrating the feasibility of large-scale automated data curation.

\item \textbf{Cost-Effective Quality Assessment}: SPICE~\citep{spice2024} enables automated difficulty assessment at 19,000$\times$ lower cost than manual labeling while maintaining high accuracy, making large-scale data curation economically viable.

\item \textbf{Practical Insights}: We provide concrete lessons learned about data quality, training strategies, and infrastructure considerations that can guide future research and development efforts.
\end{enumerate}

\subsection{Future Directions}

Looking ahead, we are extending RepoForge to support broader language ecosystems, integrating multi-agent planning for long-horizon tasks, and exploring continuous learning loops where deployed agents feed back fresh, validated data into the pipeline.

Specific areas of future work include:

\begin{itemize}
\item \textbf{Multi-Language Support}: Expanding the pipeline to handle diverse programming languages and build systems beyond Python.
\item \textbf{Continuous Learning}: Developing systems that can continuously improve from deployment feedback and new real-world examples.
\item \textbf{Long-Horizon Planning}: Integrating more sophisticated planning algorithms for complex, multi-step software engineering tasks.
\item \textbf{Real-World Deployment}: Transitioning from benchmark performance to practical deployment in production software development environments.
\end{itemize}

\subsection{Broader Impact}

Our vision is simple but ambitious: a self-improving, always-on platform that pushes the boundaries of autonomous software engineering every single day. By demonstrating that infrastructure optimizations and automated data curation can enable smaller models to achieve competitive performance, we hope to make advanced SWE capabilities more accessible to a broader range of organizations and researchers.

The techniques and insights presented in this work have the potential to accelerate progress in automated software engineering, ultimately leading to more productive and efficient software development processes across the industry. 

\begin{acks}
We would like to thank the RepoForge team for their contributions to this work. Core contributors include Zhilong Chen and Chengzong Zhao, with project leadership from Boyuan Chen and Dayi Lin. Main contributors include Yihao Chen, Arthur Leung, Gopi Krishnan Rajbahadur, and Gustavo Oliva, under the guidance of advisor Ahmed Hassan.

We would also like to thank Rui Shu and Kim Kisub for their data preparation efforts, and Alex Yang, Aaditya Bhatia, Kirill Vasilevski, Youssef Esseddiq, and Yanruo Yang for their valuable contributions to the project.

We acknowledge the broader research community whose work on software engineering benchmarks, agent frameworks, and evaluation methodologies provided the foundation for this research. We are grateful for the open-source projects and datasets that made this work possible. 
\end{acks}

\bibliographystyle{ACM-Reference-Format}
\bibliography{main}


\begin{thebibliography}{17}


\ifx \showCODEN    \undefined \def \showCODEN     #1{\unskip}     \fi
\ifx \showDOI      \undefined \def \showDOI       #1{#1}\fi
\ifx \showISBNx    \undefined \def \showISBNx     #1{\unskip}     \fi
\ifx \showISBNxiii \undefined \def \showISBNxiii  #1{\unskip}     \fi
\ifx \showISSN     \undefined \def \showISSN      #1{\unskip}     \fi
\ifx \showLCCN     \undefined \def \showLCCN      #1{\unskip}     \fi
\ifx \shownote     \undefined \def \shownote      #1{#1}          \fi
\ifx \showarticletitle \undefined \def \showarticletitle #1{#1}   \fi
\ifx \showURL      \undefined \def \showURL       {\relax}        \fi
\providecommand\bibfield[2]{#2}
\providecommand\bibinfo[2]{#2}
\providecommand\natexlab[1]{#1}
\providecommand\showeprint[2][]{arXiv:#2}

\bibitem[Bhatia et~al\mbox{.}(2025)]%
        {spice2024}
\bibfield{author}{\bibinfo{person}{Aaditya Bhatia}, \bibinfo{person}{Gustavo~A Oliva}, \bibinfo{person}{Gopi~Krishnan Rajbahadur}, \bibinfo{person}{Haoxiang Zhang}, \bibinfo{person}{Yihao Chen}, \bibinfo{person}{Zhilong Chen}, \bibinfo{person}{Arthur Leung}, \bibinfo{person}{Dayi Lin}, \bibinfo{person}{Boyuan Chen}, {and} \bibinfo{person}{Ahmed~E Hassan}.} \bibinfo{year}{2025}\natexlab{}.
\newblock \bibinfo{title}{SPICE: An Automated SWE-Bench Labeling Pipeline for Issue Clarity, Test Coverage, and Effort Estimation}.
\newblock
\newblock


\bibitem[Cao et~al\mbox{.}(2025)]%
        {skyrl2024}
\bibfield{author}{\bibinfo{person}{Shiyi Cao}, \bibinfo{person}{Sumanth Hegde}, \bibinfo{person}{Dacheng Li}, \bibinfo{person}{Tyler Griggs}, \bibinfo{person}{Shu Liu}, \bibinfo{person}{Eric Tang}, \bibinfo{person}{Jiayi Pan}, \bibinfo{person}{Xingyao Wang}, \bibinfo{person}{Akshay Malik}, \bibinfo{person}{Graham Neubig}, \bibinfo{person}{Kourosh Hakhamaneshi}, \bibinfo{person}{Richard Liaw}, \bibinfo{person}{Philipp Moritz}, \bibinfo{person}{Matei Zaharia}, \bibinfo{person}{Joseph~E. Gonzalez}, {and} \bibinfo{person}{Ion Stoica}.} \bibinfo{year}{2025}\natexlab{}.
\newblock \bibinfo{title}{SkyRL-v0: Train Real-World Long-Horizon Agents via Reinforcement Learning}.
\newblock
\newblock


\bibitem[Chowdhury et~al\mbox{.}(2024)]%
        {swebench_verified2024}
\bibfield{author}{\bibinfo{person}{Neil Chowdhury}, \bibinfo{person}{James Aung}, \bibinfo{person}{Chan~Jun Shern}, \bibinfo{person}{Oliver Jaffe}, \bibinfo{person}{Dane Sherburn}, \bibinfo{person}{Giulio Starace}, \bibinfo{person}{Evan Mays}, \bibinfo{person}{Rachel Dias}, \bibinfo{person}{Marwan Aljubeh}, \bibinfo{person}{Mia Glaese}, \bibinfo{person}{Carlos~E. Jimenez}, \bibinfo{person}{John Yang}, \bibinfo{person}{Leyton Ho}, \bibinfo{person}{Tejal Patwardhan}, \bibinfo{person}{Kevin Liu}, {and} \bibinfo{person}{Aleksander Madry}.} \bibinfo{year}{2024}\natexlab{}.
\newblock \bibinfo{title}{Introducing SWE-bench Verified}.
\newblock
\newblock
\urldef\tempurl%
\url{https://openai.com/index/introducing-swe-bench-verified/}
\showURL{%
\tempurl}


\bibitem[Jain et~al\mbox{.}(2025)]%
        {r2egym2024}
\bibfield{author}{\bibinfo{person}{Naman Jain}, \bibinfo{person}{Jaskirat Singh}, \bibinfo{person}{Manish Shetty}, \bibinfo{person}{Liang Zheng}, \bibinfo{person}{Koushik Sen}, {and} \bibinfo{person}{Ion Stoica}.} \bibinfo{year}{2025}\natexlab{}.
\newblock \showarticletitle{R2e-gym: Procedural environments and hybrid verifiers for scaling open-weights swe agents}.
\newblock \bibinfo{journal}{\emph{arXiv preprint arXiv:2504.07164}} (\bibinfo{year}{2025}).
\newblock


\bibitem[Jimenez et~al\mbox{.}(2024)]%
        {jimenez2024swebench}
\bibfield{author}{\bibinfo{person}{Carlos~E Jimenez}, \bibinfo{person}{John Yang}, \bibinfo{person}{Alexander Wettig}, \bibinfo{person}{Shunyu Yao}, \bibinfo{person}{Kexin Pei}, \bibinfo{person}{Ofir Press}, {and} \bibinfo{person}{Karthik~R Narasimhan}.} \bibinfo{year}{2024}\natexlab{}.
\newblock \showarticletitle{{SWE}-bench: Can Language Models Resolve Real-world Github Issues?}. In \bibinfo{booktitle}{\emph{The Twelfth International Conference on Learning Representations}}.
\newblock
\urldef\tempurl%
\url{https://openreview.net/forum?id=VTF8yNQM66}
\showURL{%
\tempurl}


\bibitem[Jin et~al\mbox{.}(2025)]%
        {searchr1_2024}
\bibfield{author}{\bibinfo{person}{Bowen Jin}, \bibinfo{person}{Hansi Zeng}, \bibinfo{person}{Zhenrui Yue}, \bibinfo{person}{Jinsung Yoon}, \bibinfo{person}{Sercan Arik}, \bibinfo{person}{Dong Wang}, \bibinfo{person}{Hamed Zamani}, {and} \bibinfo{person}{Jiawei Han}.} \bibinfo{year}{2025}\natexlab{}.
\newblock \showarticletitle{Search-r1: Training llms to reason and leverage search engines with reinforcement learning}.
\newblock \bibinfo{journal}{\emph{arXiv preprint arXiv:2503.09516}} (\bibinfo{year}{2025}).
\newblock


\bibitem[Li et~al\mbox{.}(2025)]%
        {torl2024}
\bibfield{author}{\bibinfo{person}{Xuefeng Li}, \bibinfo{person}{Haoyang Zou}, {and} \bibinfo{person}{Pengfei Liu}.} \bibinfo{year}{2025}\natexlab{}.
\newblock \bibinfo{title}{ToRL: Scaling Tool-Integrated RL}.
\newblock
\newblock
\showeprint[arxiv]{2503.23383}~[cs.CL]
\urldef\tempurl%
\url{https://arxiv.org/abs/2503.23383}
\showURL{%
\tempurl}


\bibitem[Moritz et~al\mbox{.}(2018)]%
        {ray2018}
\bibfield{author}{\bibinfo{person}{Philipp Moritz}, \bibinfo{person}{Robert Nishihara}, \bibinfo{person}{Stephanie Wang}, \bibinfo{person}{Alexey Tumanov}, \bibinfo{person}{Richard Liaw}, \bibinfo{person}{Eric Liang}, \bibinfo{person}{Melih Elibol}, \bibinfo{person}{Zongheng Yang}, \bibinfo{person}{William Paul}, \bibinfo{person}{Michael~I Jordan}, {et~al\mbox{.}}} \bibinfo{year}{2018}\natexlab{}.
\newblock \showarticletitle{Ray: A distributed framework for emerging AI applications}.
\newblock \bibinfo{journal}{\emph{Proceedings of the 13th USENIX symposium on operating systems design and implementation}} (\bibinfo{year}{2018}), \bibinfo{pages}{561--577}.
\newblock


\bibitem[Pan et~al\mbox{.}(2024)]%
        {swegym2024}
\bibfield{author}{\bibinfo{person}{Jiayi Pan}, \bibinfo{person}{Xingyao Wang}, \bibinfo{person}{Graham Neubig}, \bibinfo{person}{Navdeep Jaitly}, \bibinfo{person}{Heng Ji}, \bibinfo{person}{Alane Suhr}, {and} \bibinfo{person}{Yizhe Zhang}.} \bibinfo{year}{2024}\natexlab{}.
\newblock \bibinfo{title}{Training Software Engineering Agents and Verifiers with SWE-Gym}.
\newblock
\newblock
\showeprint[arxiv]{2412.21139}~[cs.SE]
\urldef\tempurl%
\url{https://arxiv.org/abs/2412.21139}
\showURL{%
\tempurl}


\bibitem[Sheng et~al\mbox{.}(2024)]%
        {verl2024}
\bibfield{author}{\bibinfo{person}{Guangming Sheng}, \bibinfo{person}{Chi Zhang}, \bibinfo{person}{Zilingfeng Ye}, \bibinfo{person}{Xibin Wu}, \bibinfo{person}{Wang Zhang}, \bibinfo{person}{Ru Zhang}, \bibinfo{person}{Yanghua Peng}, \bibinfo{person}{Haibin Lin}, {and} \bibinfo{person}{Chuan Wu}.} \bibinfo{year}{2024}\natexlab{}.
\newblock \showarticletitle{HybridFlow: A Flexible and Efficient RLHF Framework}.
\newblock \bibinfo{journal}{\emph{arXiv preprint arXiv: 2409.19256}} (\bibinfo{year}{2024}).
\newblock


\bibitem[Sun et~al\mbox{.}(2025)]%
        {swefactory2024}
\bibfield{author}{\bibinfo{person}{Haoran Sun}, \bibinfo{person}{Haoyu Bian}, \bibinfo{person}{Shaoning Zeng}, \bibinfo{person}{Yunbo Rao}, \bibinfo{person}{Xu Xu}, \bibinfo{person}{Lin Mei}, {and} \bibinfo{person}{Jianping Gou}.} \bibinfo{year}{2025}\natexlab{}.
\newblock \showarticletitle{DatasetAgent: A Novel Multi-Agent System for Auto-Constructing Datasets from Real-World Images}.
\newblock \bibinfo{journal}{\emph{arXiv preprint arXiv:2507.08648}} (\bibinfo{year}{2025}).
\newblock


\bibitem[Tan et~al\mbox{.}(2025)]%
        {rllm2025}
\bibfield{author}{\bibinfo{person}{Sijun Tan}, \bibinfo{person}{Michael Luo}, \bibinfo{person}{Colin Cai}, \bibinfo{person}{Tarun Venkat}, \bibinfo{person}{Kyle Montgomery}, \bibinfo{person}{Aaron Hao}, \bibinfo{person}{Tianhao Wu}, \bibinfo{person}{Arnav Balyan}, \bibinfo{person}{Manan Roongta}, \bibinfo{person}{Chenguang Wang}, \bibinfo{person}{Li~Erran Li}, \bibinfo{person}{Raluca~Ada Popa}, {and} \bibinfo{person}{Ion Stoica}.} \bibinfo{year}{2025}\natexlab{}.
\newblock \bibinfo{title}{rLLM: A Framework for Post-Training Language Agents}.
\newblock \bibinfo{howpublished}{\url{https://pretty-radio-b75.notion.site/rLLM-A-Framework-for-Post-Training-Language-Agents-21b81902c146819db63cd98a54ba5f31}}.
\newblock
\newblock
\shownote{Notion Blog}.


\bibitem[Wang(2025)]%
        {openhands_lm_32b2024}
\bibfield{author}{\bibinfo{person}{Xingyao Wang}.} \bibinfo{year}{2025}\natexlab{}.
\newblock \showarticletitle{Introducing OpenHands LM 32B -- A Strong, Open Coding Agent Model}.
\newblock \bibinfo{journal}{\emph{All Hands AI Blog}} (\bibinfo{date}{31 March} \bibinfo{year}{2025}).
\newblock
\urldef\tempurl%
\url{https://www.all-hands.dev/blog/introducing-openhands-lm-32b----a-strong-open-coding-agent-model}
\showURL{%
\tempurl}


\bibitem[Wang et~al\mbox{.}(2025)]%
        {openhands2024}
\bibfield{author}{\bibinfo{person}{Xingyao Wang}, \bibinfo{person}{Boxuan Li}, \bibinfo{person}{Yufan Song}, \bibinfo{person}{Frank~F. Xu}, \bibinfo{person}{Xiangru Tang}, \bibinfo{person}{Mingchen Zhuge}, \bibinfo{person}{Jiayi Pan}, \bibinfo{person}{Yueqi Song}, \bibinfo{person}{Bowen Li}, \bibinfo{person}{Jaskirat Singh}, \bibinfo{person}{Hoang~H. Tran}, \bibinfo{person}{Fuqiang Li}, \bibinfo{person}{Ren Ma}, \bibinfo{person}{Mingzhang Zheng}, \bibinfo{person}{Bill Qian}, \bibinfo{person}{Yanjun Shao}, \bibinfo{person}{Niklas Muennighoff}, \bibinfo{person}{Yizhe Zhang}, \bibinfo{person}{Binyuan Hui}, \bibinfo{person}{Junyang Lin}, \bibinfo{person}{Robert Brennan}, \bibinfo{person}{Hao Peng}, \bibinfo{person}{Heng Ji}, {and} \bibinfo{person}{Graham Neubig}.} \bibinfo{year}{2025}\natexlab{}.
\newblock \showarticletitle{OpenHands: An Open Platform for {AI} Software Developers as Generalist Agents}. In \bibinfo{booktitle}{\emph{The Thirteenth International Conference on Learning Representations}}.
\newblock
\urldef\tempurl%
\url{https://openreview.net/forum?id=OJd3ayDDoF}
\showURL{%
\tempurl}


\bibitem[Wu et~al\mbox{.}(2025)]%
        {mmsearch_r1_2024}
\bibfield{author}{\bibinfo{person}{Jinming Wu}, \bibinfo{person}{Zihao Deng}, \bibinfo{person}{Wei Li}, \bibinfo{person}{Yiding Liu}, \bibinfo{person}{Bo You}, \bibinfo{person}{Bo Li}, \bibinfo{person}{Zejun Ma}, {and} \bibinfo{person}{Ziwei Liu}.} \bibinfo{year}{2025}\natexlab{}.
\newblock \showarticletitle{MMSearch-R1: Incentivizing LMMs to Search}.
\newblock \bibinfo{journal}{\emph{arXiv preprint arXiv:2506.20670}} (\bibinfo{year}{2025}).
\newblock


\bibitem[Zeng et~al\mbox{.}(2025)]%
        {skywork2024swe}
\bibfield{author}{\bibinfo{person}{Liang Zeng}, \bibinfo{person}{Yongcong Li}, \bibinfo{person}{Yuzhen Xiao}, \bibinfo{person}{Changshi Li}, \bibinfo{person}{Chris~Yuhao Liu}, \bibinfo{person}{Rui Yan}, \bibinfo{person}{Tianwen Wei}, \bibinfo{person}{Jujie He}, \bibinfo{person}{Xuchen Song}, \bibinfo{person}{Yang Liu}, {et~al\mbox{.}}} \bibinfo{year}{2025}\natexlab{}.
\newblock \showarticletitle{Skywork-SWE: Unveiling Data Scaling Laws for Software Engineering in LLMs}.
\newblock \bibinfo{journal}{\emph{arXiv preprint arXiv:2506.19290}} (\bibinfo{year}{2025}).
\newblock


\bibitem[Zhang et~al\mbox{.}(2025)]%
        {swebench_live2024}
\bibfield{author}{\bibinfo{person}{Linghao Zhang}, \bibinfo{person}{Shilin He}, \bibinfo{person}{Chaoyun Zhang}, \bibinfo{person}{Yu Kang}, \bibinfo{person}{Bowen Li}, \bibinfo{person}{Chengxing Xie}, \bibinfo{person}{Junhao Wang}, \bibinfo{person}{Maoquan Wang}, \bibinfo{person}{Yufan Huang}, \bibinfo{person}{Shengyu Fu}, \bibinfo{person}{Elsie Nallipogu}, \bibinfo{person}{Qingwei Lin}, \bibinfo{person}{Yingnong Dang}, \bibinfo{person}{Saravan Rajmohan}, {and} \bibinfo{person}{Dongmei Zhang}.} \bibinfo{year}{2025}\natexlab{}.
\newblock \showarticletitle{SWE-bench Goes Live!}
\newblock \bibinfo{journal}{\emph{arXiv preprint arXiv:2505.23419}} (\bibinfo{year}{2025}).
\newblock


\end{thebibliography}

\end{document}